\documentclass[manuscript]{acmart}

\usepackage[ruled, vlined, linesnumbered]{algorithm2e}
\usepackage{amsfonts}
\usepackage{subcaption}
\usepackage{multirow}
\usepackage{bm}
\usepackage{bbold}
\usepackage{mathtools}
\usepackage[utf8]{inputenc}
\usepackage{graphicx}
\usepackage{booktabs}
\DeclareUnicodeCharacter{2212}{−}

\newcommand{\xhdr}[1]{\vspace{0.8mm}\noindent{{\bf #1}}}

\AtBeginDocument{%
  \providecommand\BibTeX{{%
    \normalfont B\kern-0.5em{\scshape i\kern-0.25em b}\kern-0.8em\TeX}}}

\copyrightyear{2022} 
\acmYear{2022} 
\setcopyright{acmlicensed}\acmConference[RecSys '22]{Sixteenth ACM Conference on Recommender Systems}{September 18--23, 2022}{Seattle, WA, USA}
\acmBooktitle{Sixteenth ACM Conference on Recommender Systems (RecSys '22), September 18--23, 2022, Seattle, WA, USA}
\acmPrice{15.00}
\acmDOI{10.1145/3523227.3546770}
\acmISBN{978-1-4503-9278-5/22/09}

\begin{document}

\title{Defending Substitution-Based Profile Pollution Attacks on Sequential Recommenders}


\author{Zhenrui Yue}
\affiliation{%
  \institution{University of Illinois Urbana-Champaign}
  \country{USA}
}
\email{zhenrui3@illinois.edu}

\author{Huimin Zeng}
\affiliation{%
  \institution{University of Illinois Urbana-Champaign}
  \country{USA}
}
\email{huiminz3@illinois.edu}

\author{Ziyi Kou}
\affiliation{%
  \institution{University of Illinois Urbana-Champaign}
  \country{USA}
}
\email{ziyikou2@illinois.edu}

\author{Lanyu Shang}
\affiliation{%
  \institution{University of Illinois Urbana-Champaign}
  \country{USA}
}
\email{lshang3@illinois.edu}

\author{Dong Wang}
\affiliation{%
  \institution{University of Illinois Urbana-Champaign}
  \country{USA}
}
\email{dwang24@illinois.edu}

\renewcommand{\shortauthors}{Yue, et al.}

\begin{abstract}
While sequential recommender systems achieve significant improvements on capturing user dynamics, we argue that sequential recommenders are vulnerable against substitution-based profile pollution attacks. To demonstrate our hypothesis, we propose a substitution-based adversarial attack algorithm, which modifies the input sequence by selecting certain vulnerable elements and substituting them with adversarial items. In both untargeted and targeted attack scenarios, we observe significant performance deterioration using the proposed profile pollution algorithm. Motivated by such observations, we design an efficient adversarial defense method called Dirichlet neighborhood sampling. Specifically, we sample item embeddings from a convex hull constructed by multi-hop neighbors to replace the original items in input sequences. During sampling, a Dirichlet distribution is used to approximate the probability distribution in the neighborhood such that the recommender learns to combat local perturbations. Additionally, we design an adversarial training method tailored for sequential recommender systems. In particular, we represent selected items with one-hot encodings and perform gradient ascent on the encodings to search for the worst case linear combination of item embeddings in training. As such, the embedding function learns robust item representations and the trained recommender is resistant to test-time adversarial examples. Extensive experiments show the effectiveness of both our attack and defense methods, which consistently outperform baselines by a significant margin across model architectures and datasets.
\end{abstract}



\maketitle


\section{Introduction}

Sequential recommenders are a common approach for making personalized recommendations, such recommenders take user interaction history as input and generate potential items that may be of interest to the user~\cite{li2017neural, kang2018self,zhang2019through}. Different from traditional recommenders, sequential recommendation can capture users' evolving dynamics by treating item transition patterns as temporal sequences. As such, various recent sequential recommenders (e.g., Locker\cite{he2021locker}) consistently outperform previous state-of-the-art models. 

Nevertheless, the reliability of recommenders is known to deteriorate under adversarial perturbations~\cite{goodfellow2014explaining, zhang2018crowdsourcing, yue2021black}. While gradient-based methods can not be directly applied on recommenders for searching perturbations in the item space, certain heuristics can be used to perform adversarial attacks~\cite{o2005recommender, yang2017fake}. Based on such methods, profile pollution~\footnote{In the following, we use adversarial attack and profile pollution attack interchangeably.} is adopted to perform user-specific adversarial attacks~\cite{yang2017fake, yue2021black}. To perform such attacks, malware can be used to access user profiles and interactions (e.g., cross-site request forgeries (CSRF) or backdoor attacks)~\cite{zeller2008cross, ren2015towards, lee2017all}.

We argue that sequential recommenders have unique vulnerabilities against substitution-based profile pollution attacks. Substitution-based adversarial attacks manipulate vulnerable items in previous or ongoing interactions
to sabotage the recommendation (i.e., untargeted attacks) or manipulate the recommended items (i.e., targeted attacks)~\cite{zeller2008cross, ren2015towards, lee2017all, marshall2016mood}.
Previous works on profile pollution study attack algorithms and have the following limitations: (1)~Existing methods designed for traditional recommenders can not be applied, or are not tailored for sequential recommenders~\cite{yang2017fake, tang2020revisiting, zhang2019understanding}; (2)~Previous methods do not explore substitution-based attacks, instead, they focus on the injection of adversarial items~\cite{zhang2019understanding, yue2021black}; and (3)~Adversarial defense methods are not discussed~\cite{yang2017fake, zhang2019understanding, tang2020revisiting, yue2021black, oh2022robustness}.

A similar sequential setting can be found in natural language processing (NLP)~\cite{ebrahimi2017hotflip}. NLP adversarial attacks primarily study word-level substitutions, where a small percentage of input words are substituted~\cite{alzantot2018generating, li2018textbugger, ren2019generating}. To find adversarial examples in NLP, one possible method is to search among synonyms iteratively~\cite{alzantot2018generating}. Another common approach is to select vulnerable words and replace them with adversarial ones~\cite{li2018textbugger, ren2019generating}. Despite similarity between NLP and sequential recommendation, it is difficult to build adversarial examples for recommenders using similar techniques. The reasons are two-fold: (1) the item space is too large to enumerate without synonym information in recommender systems; (2) existing NLP attack algorithms do not exploit gradient-based methods to maximize the attack performance.

In this paper, we design a gradient-guided algorithm for substitution-based profile pollution attacks. Our attack algorithm modifies the input sequences by selecting a limited number of vulnerable elements and replacing them with carefully chosen adversarial items. By design, the substitution-based attack can perform either substitution attacks in a sequence, or inject malicious items to user interactions (by inserting and attacking an item, as in~\cite{yue2021black}). To prevent from being identified as malicious input, we impose two constraints:
(1)~we restrict the maximum number of substitutions in each sequence; (2)~for substitute items, we additionally require similarity to the original items.
Our experiment results demonstrate a significant performance deterioration in both untargeted and targeted scenarios, suggesting that the proposed substitution-based attacks pose critical threats to sequential recommenders.

Motivated by the research gap in defense methods, we additionally propose two defense methods that are tailored for sequential recommendation. Dirichlet neighborhood sampling is proposed for efficient training of robust recommenders. Specifically, we construct multi-hop neighborhoods with the embedding matrix and sample augmented item embeddings via Dirichlet distributions to perform random substitutions. Additionally, we design an adversarial training method for sequential recommenders. In particular, we represent items with one-hot encodings and update the encodings to search for the worst case augmentation.
By training the recommender on such virtual input, we 
extend the robustness of the recommender to a larger neighborhood. 
Our evaluation results show that the proposed defense methods significantly reduce performance variations under profile pollution attacks by over $50\%$ on multiple real-world datasets.

\section{Related Work}

\subsection{Adversarial Attacks in Recommender Systems}
Adversarial attacks on recommender systems can be categorized into two different classes: (1)~\emph{data poisoning} and (2)~\emph{profile pollution}. Data poisoning attacks craft fake user profiles and inject such profiles into the training data,
causing biased recommendation results upon deployment~\cite{christakopoulou2019adversarial, song2020poisonrec, fang2020influence, tang2020revisiting, zhang2020practical, huang2021data}. Unlike poisoning attacks, profile pollution attacks alter existing user profiles (i.e., user interactions) to manipulate recommended items~\cite{xing2013take, meng2014your, yang2017fake, zhang2019understanding, yue2021black}. Profile pollution can be performed via security breaches like web injection, CSRF attacks or malware~\cite{zeller2008cross, ren2015towards, lee2017all, zhang2019understanding}. 

In this work, we focus on profile pollution and perform user-specific attacks to manipulate the recommendation results~\cite{yue2021black}. CSRF attacks manipulate user history on websites like YouTube under a black-box setting~\cite{xing2013take}. Malicious requests are exploited to bias web advertisements towards higher-paying advertisers~\cite{meng2014your}. With knowledge to the recommender, finding adversarial examples can be formulated as an optimization problem~\cite{yang2017fake}. Web injections are used in unprotected websites to promote target items~\cite{zhang2019understanding}. \citet{yue2021black} propose to extract black-box recommenders and compute adversarial examples using the extracted model. Our proposed profile pollution algorithm differs from the previous methods in two aspects: (1) we extend the attack scope to substitutions in the full sequence; and (2) constraints are imposed on adversarial examples to enforce similarity between the clean and adversarial sequences.


\subsection{Adversarial Attacks in Language}
Given the setting of profile pollution by item substitution, a similar scenario can be found on various NLP classification tasks, where adversarial attacks are studied on word-level substitutions. NLP substitution attacks replace a small percentage of the input words with adversarial words to achieve the attack objective~\cite{alzantot2018generating, li2018textbugger}. NLP adversarial examples can be generated using a genetic algorithm, where synonyms are iterated and evaluated~\cite{alzantot2018generating}. The vulnerability of input words can be evaluated via importance scores, followed by substituting the most vulnerable words to construct adversarial examples until the model is fooled~\cite{li2018textbugger, ren2019generating, li2020bert, garg2020bae}. 
Despite similarity between language and sequential recommendation, it is difficult to transfer existing NLP methods to recommenders due to: (1)~the lack of synonyms (i.e., neighbor items) in recommendation data; and (2)~the inefficiency of existing NLP methods for large-scale or real-time attacks. Unlike previous methods, the proposed profile pollution algorithm constructs multi-hop neighborhoods and efficiently computes adversarial substitutes by exploiting gradient information from the recommender model.

\subsection{Adversarial Defense Methods}
Various defense methods are proposed to enhance the robustness of neural networks~\cite{goodfellow2014explaining, gowal2018effectiveness}. Adversarial training and smoothing methods incorporate augmented examples in training to enhance robustness against perturbations~\cite{goodfellow2014explaining, zeng2021adversarial, zeng2021certified}. Interval bound propagation regularizes tractable upper and lower bounds to guarantee robustness~\cite{gowal2018effectiveness}. Adversarial training introduced norm-bounded local perturbations to word embeddings to learn robust features~\cite{sato2018interpretable, zhu2019freelb}. Recently, Dirichlet neighborhood ensemble leverages synonym embeddings to build robust word representations~\cite{zhou2020defense}. For recommender systems, adversarial training introduces perturbations to improve model performance and generalization~\cite{he2018adversarial, yuan2019adversarial, chen2019adversarial}. 
Multimedia recommender learns to defend perturbations on product images via adversarial training~\cite{tang2019adversarial}. To the best of our knowledge, no previous methods attempt to enhance the robustness of recommender systems against profile pollution attacks. As such, we propose two novel defense methods tailored for sequential recommenders and show that model robustness can be significantly improved using the proposed defense methods.

\section{Preliminaries}

\subsection{Setting}
\subsubsection{Data} 
Our framework is based on sequential input, which takes user interaction history $\bm{x}$ (sorted by timestamps) as input. $\bm{x}$ represents input $[x_1, x_2, ..., x_l]$ of length $l$, with each element represented in the item scope $\mathcal{I}$ (i.e., $x_i \in \mathcal{I}$). The next item interaction $x_{l+1} \in \mathcal{I}$ after input $\bm{x}$ is used as ground truth $y$ (i.e., $y = x_{l+1}$).

\subsubsection{Model} 
We denote the sequential recommender with function $\bm{f}$. Given input sequence $\bm{x}$, $\bm{f}$ predicts a probability distribution over the item scope $\mathcal{I}$. $\bm{f}$ comprises of an embedding function $\bm{f}_e$ and a sequential model $\bm{f}_m$, with $\bm{f}(\bm{x}) = \bm{f}_m(\bm{f}_e(\bm{x}))$. For data pair $(\bm{x}, y)$, ideally, $\bm{f}$ predicts $y$ with the highest probability (i.e., $y = \arg\max\bm{f}(\bm{x})$).

\subsubsection{Optimization} 
The learning of a sequential recommender $\bm{f}$ is to maximize the probability of output item $y$ upon input $\bm{x}$. In other words, we minimize the expectation of loss $\mathcal{L}$ w.r.t. $\bm{f}$ over data distribution $\mathcal{X}$:
\begin{equation}
    \min_{\substack{\bm{f}}} \mathbb{E}_{(\bm{x}, y) \sim \mathcal{X}} \mathcal{L}(\bm{f}(\bm{x}), y),
\end{equation}
where $\mathcal{L}$ represents the training loss function (i.e., ranking loss or cross entropy loss).

\subsection{Adversarial Attacks and Defense}

\subsubsection{Assumptions} 
We first introduce the assumptions of our framework in the attack scenario, we formalize the problem setting with the following assumptions to define our research scope:
\begin{itemize}
    \item \emph{Data Access}: Evaluation data $\mathcal{X}$ and item scope $\mathcal{I}$ are accessible to the attacker to perform the attacks.
    \item \emph{White-box Attacks}: We assume weights of $\bm{f}$ can be accessed by the attacker to compute adversarial examples. 
    \item \emph{Limited Substitutions}: In each input sequence, we impose a maximum number of substitutions $z$ to enforce the similarity between the original and adversarial examples on the sequence level.
    \item \emph{Constraints on Item Similarity}: To avoid being identified as outlier, we impose a similarity constraint with a minimum cosine similarity $\tau$ between the original items and the adversarial items in the embedding space.
\end{itemize}
The above assumptions are made for a direct and undisturbed evaluation of the proposed profile pollution attack algorithm (i.e., without considering external influences like model extraction). The constructed adversarial attacks are also more challenging and thus, evaluate the effectiveness of the proposed defense methods under harsh conditions. For the defense propose, we assume full access to the data and recommender models.

\subsubsection{Threat Model}
As described, we adopt profile pollution attacks to construct adversarial examples. We investigate untargeted attacks (i.e., demotion) and targeted attacks (i.e., promotion). 
We formulate both attacks below.
\begin{itemize}
    \item \emph{Untargeted Attack}: 
    Untargeted adversarial examples are constructed to minimize the probability w.r.t. ground truth $y$. From the assumptions we require: (1)~The number of substitutions is constrained by Hamming distance (i.e., at most $z$ substitutions); (2)~Minimum cosine similarity $\tau$ is required between original and adversarial items. We denote the adversarial example with $\bm{x}'$ and formulate the attack as an optimization problem w.r.t. $\bm{x}$:
    \begin{equation}
        \bm{x}' = \arg\max_{\substack{\bm{x}}} \mathcal{L}(\bm{f}(\bm{x}), y) \quad s.t. \quad \mathrm{Hamming}(\bm{x}, \bm{x}') \leq z, \quad \mathrm{CosSim}(x_i, x'_i) \geq \tau, \quad i = 1, 2, \ldots
    \end{equation}
    \item \emph{Targeted Attack:} Targeted adversarial examples aim to maximize the probability w.r.t. target item $t$. Specifically, we minimize the loss w.r.t. input $\bm{x}$. Similar constraints are imposed for targeted attacks:
    \begin{equation}
        \bm{x}' = \arg\min_{\substack{\bm{x}}} \mathcal{L}(\bm{f}(\bm{x}), t) \quad s.t. \quad \mathrm{Hamming}(\bm{x}, \bm{x}') \leq z, \quad \mathrm{CosSim}(x_i, x'_i) \geq \tau, \quad i = 1, 2, \ldots
    \end{equation}
\end{itemize}

\subsubsection{Adversarial Training} 
Given the profile pollution attack framework, a robust recommender can be built by minimizing the expected loss upon perturbed input data (i.e., adversarial training). That is, we alternate between optimizing model parameters in $\bm{f}$ and updating adversarial data based on $\bm{x}$ to search for an equilibrium. The adversarial examples are generated on the fly during training and being used to minimize the training loss $\mathcal{L}$, such that the model $\bm{f}$ is forced to adjust parameters to resist adversarial perturbations. Formally, we define the adversarial training as a minimax optimization problem w.r.t. model $\bm{f}$:
\begin{equation}
    \min_{\substack{\bm{f}}} \mathbb{E}_{(\bm{x}, y) \sim \mathcal{X}} \left[ \max_{\substack{\bm{x}}} \mathcal{L}(\bm{f}(\bm{x}), y) \right].
\end{equation}
\section{Methodology}

\begin{figure*}[t]
    \centering
    \includegraphics[trim=0 7.5cm 7.5cm 0, clip, width=0.8\linewidth]{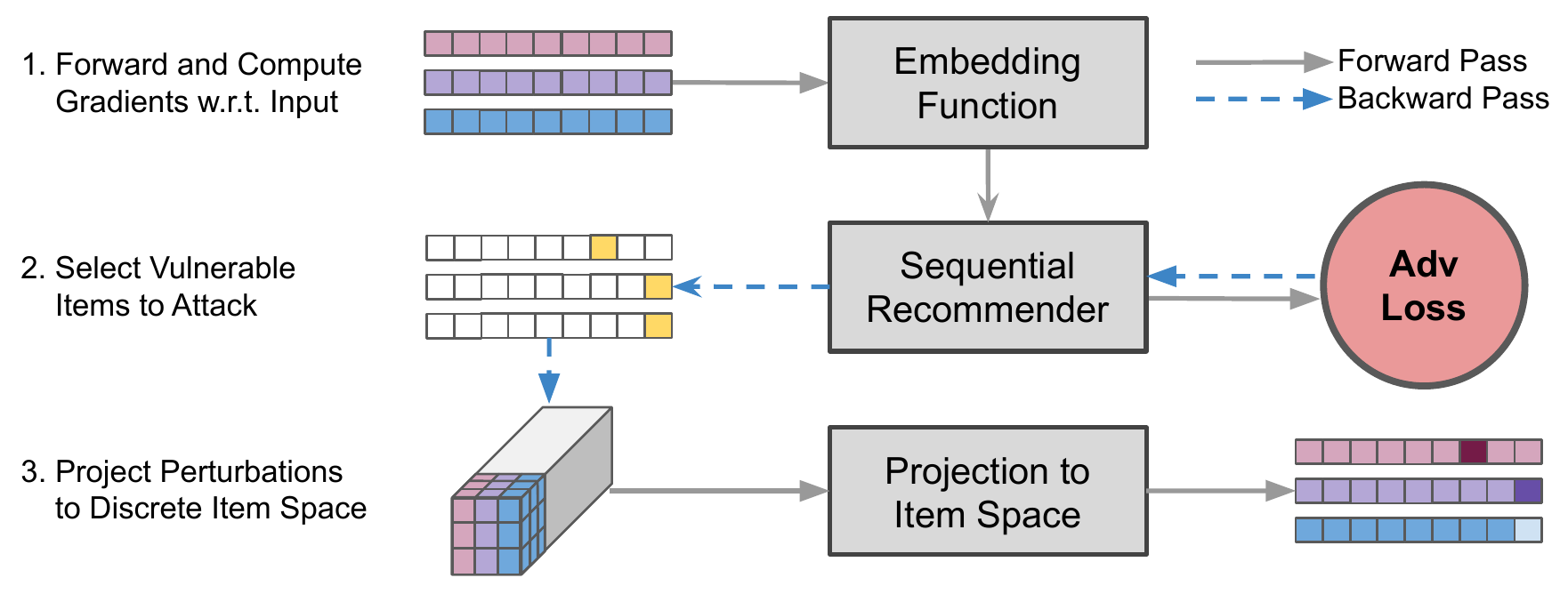}
    \caption{The proposed profile pollution attack. We first forward clean input sequences and compute gradients w.r.t. the input with the adversarial loss. Then, we compute importance scores to select vulnerable items in input sequences. In the last step, we project the continuous perturbations of the selected items to the discrete item space based on cosine similarity.} 
    \label{fig:attack}
    \vspace{-10pt}
\end{figure*}

\subsection{Profile Pollution Attack}
\label{sec:attack}
The proposed algorithm can either attack history items or insert adversarial items to `fool' recommenders. Ideally, the polluted sequences result in either performance degrade (i.e., untargeted attack) or increased exposure of the target item (i.e., targeted attack). We illustrate the proposed profile pollution attack algorithm in Figure~\ref{fig:attack}, the detailed steps for both untargeted and targeted attacks are described as in Algorithm~\ref{alg:attack}.

\begin{algorithm}[t]
    \caption{Profile Pollution Attack Algorithm}
    \label{alg:attack}
    \SetAlgoLined
    \textbf{Input}
    Sequence $\bm{x}$, target $t$, recommender $\bm{f}$ (i.e., $\bm{f}_{e}$, $\bm{f}_{m}$), max substitutions $z$, similarity threshold $\tau$\;
    \textbf{Input}
    Target $t$ if targeted attack\;
    \textbf{Output}
    Polluted sequence $\bm{x}'$\;
    \tcp{Step 1: Forward and Compute Gradients}
    Initialize polluted sequence: $\bm{x}' \leftarrow \bm{x}$\;
    Compute sequence embeddings: $\tilde{\bm{x}}' \leftarrow \bm{f}_{e}(\bm{x}')$\;
    \uIf{$\mathrm{targeted}$}{
        Compute gradients w.r.t. $\tilde{\bm{x}}'$: $\nabla_{\tilde{\bm{x}}'} \leftarrow \nabla_{\tilde{\bm{x}}'} \mathcal{L}_{\mathit{ce}} (\bm{f}_{m}(\tilde{\bm{x}}'), t)$\;
    }
    \Else{
        Compute gradients w.r.t. $\tilde{\bm{x}}'$: $\nabla_{\tilde{\bm{x}}'} \leftarrow - \nabla_{\tilde{\bm{x}}'} \mathcal{L}_{\mathit{ce}} (\bm{f}_{m}(\tilde{\bm{x}}'), \arg\max(\bm{f}_{m}(\tilde{\bm{x}}')))$\;
    }
    \tcp{Step 2: Select Vulnerable Items}
    Compute importance ranking: $\bm{r} \leftarrow \mathrm{argsort}(\| \nabla_{\tilde{\bm{x}}'} \|, \mathrm{descending})$\;
    Select $z$ most vulnerable items in $\bm{x}'$: $\bm{r} \leftarrow \bm{r}\left[ :z \right]$\;
    \tcp{Step 3: Project Attacks to Item Space}
    \For{$i \in \bm{r}$}{ 
        Compute cosine similarity $\bm{s}$: $\bm{s} \leftarrow \mathrm{CosSim}(\tilde{\bm{x}}'_i - \textit{sign}(\nabla_{\tilde{\bm{x}}'_i}), \bm{f}_{e}(c))) \forall c \in \mathcal{I}$\;
        Impose similarity constraint $\bm{s}_c$: $\bm{s}_c \leftarrow \mathbb{1}(\mathrm{CosSim}(\tilde{\bm{x}}'_i, \bm{f}_{e}(c)) \geq \tau) \forall c \in \mathcal{I}$\;
        Update cosine similarity $\bm{s}$: $\bm{s} \leftarrow \bm{s} \odot \bm{s}_c$\;
        Replace $x'_i$ in $\bm{x}'$: $x'_i \leftarrow \arg\max(\bm{s})$\;
    }
\end{algorithm}

\xhdr{Step 1: Forward and Compute Gradients w.r.t. Input.}
Given input $\bm{x}$, we construct the polluted sequence $\bm{x}'$ by substituting selected items in $\bm{x}$. First, we initialize $\bm{x}'$ unchanged to $\bm{x}$, followed by computing the embedded sequence $\tilde{\bm{x}}'$ using the embedding function $\bm{f}_e$. We then calculate the gradients of $\tilde{\bm{x}}'$ to construct adversarial perturbations in the embedding space. In the case of targeted attack, $\tilde{\bm{x}}'$ is fed through the recommender function $\bm{f}_m$ to compute the cross entropy loss $\mathcal{L}_{ce}$ w.r.t. target item $t$. Then, backward propagation is performed to retrieve the gradients $\nabla_{\tilde{\bm{x}}'}$ of $\tilde{\bm{x}}'$ (i.e. $\nabla_{\tilde{\bm{x}}'} = \nabla_{\tilde{\bm{x}}'} \mathcal{L}_{\mathit{ce}} (\bm{f}_{m}(\tilde{\bm{x}}'), t)$). For untargeted attacks, we compute the cross entropy loss w.r.t. the predicted item in the output distribution.
The gradients are negated in untargeted attacks to perform gradient ascent.

\xhdr{Step 2: Select Vulnerable Items to Attack.}
In input sequence, items are of different vulnerability. We select $z$ vulnerable items and perform substitution on such items to achieve the best attack performance. In particular, we calculate importance scores using $\nabla_{\tilde{\bm{x}}'}$ from the previous step. Unlike existing works, we take $l^2$ norm of the last dimension in $\nabla_{\tilde{\bm{x}}'}$ as our importance scores (i.e., $\| \nabla_{\tilde{\bm{x}}'} \|$), such that the vulnerability of input items can be efficiently estimated with the steepness of the gradients. We select at most $z$ items by choosing the first $z$ items from importance ranking $\bm{r}$. Based on the fast gradient sign method~\cite{goodfellow2014explaining}, the perturbed embeddings $\tilde{\bm{x}}'$ can be computed as follows:
\begin{itemize}
    \item Untargeted: 
    \begin{equation}
        \tilde{\bm{x}}' = \tilde{\bm{x}}' - \mathit{sign}(-\nabla_{\tilde{\bm{x}}'} \mathcal{L}_{ce}(\bm{f}_{m}(\tilde{\bm{x}}'), \arg\max(\bm{f}_{m}(\tilde{\bm{x}}'))))
    \end{equation}
    \item Targeted: 
    \begin{equation}
        \tilde{\bm{x}}' = \tilde{\bm{x}}' - \mathit{sign}(\nabla_{\tilde{\bm{x}}'} \mathcal{L}_{ce}(\bm{f}_{m}(\tilde{\bm{x}}'), t)),
    \end{equation}
\end{itemize}

\xhdr{Step 3: Project Attacks to Item Space.}
In the last step, we project the perturbed embeddings $\tilde{\bm{x}}'$ back to the item space. Similar to~\cite{yue2021black}, we compute cosine similarity between $\tilde{\bm{x}}'$ and candidate items from $\mathcal{I}$, items with higher similarity values are favored as adversarial substitutes. The idea behind it is to find closest items in the direction of 
the constructed adversarial embeddings using cosine similarity.
We denote $\bm{s}$ as the cosine similarity scores. Here, we impose another constraint to enforce item similarity, where adversarial items are required to have a minimum cosine similarity (above $\tau$) with the original items. We use the indicator function (i.e., $\mathbb{1}$) to select items fulfilling the similarity constraint. $\bm{s}_c$ denotes the similarity constraint results and $\bm{s}_c$ is element-wise multiplied (i.e., $\odot$) with $\bm{s}$. The final adversarial can be selected from $\bm{s}$ by taking the highest similarity score. To improve the efficiency of the adversarial attack and item projection, the computation of all previous steps is performed batch-wise with matrix multiplication.

\begin{figure*}[t]
  \centering
  \begin{subfigure}{0.49\textwidth}
  \centering
    \includegraphics[trim=5cm 2cm 6cm 4cm, clip, width=1.0\linewidth]{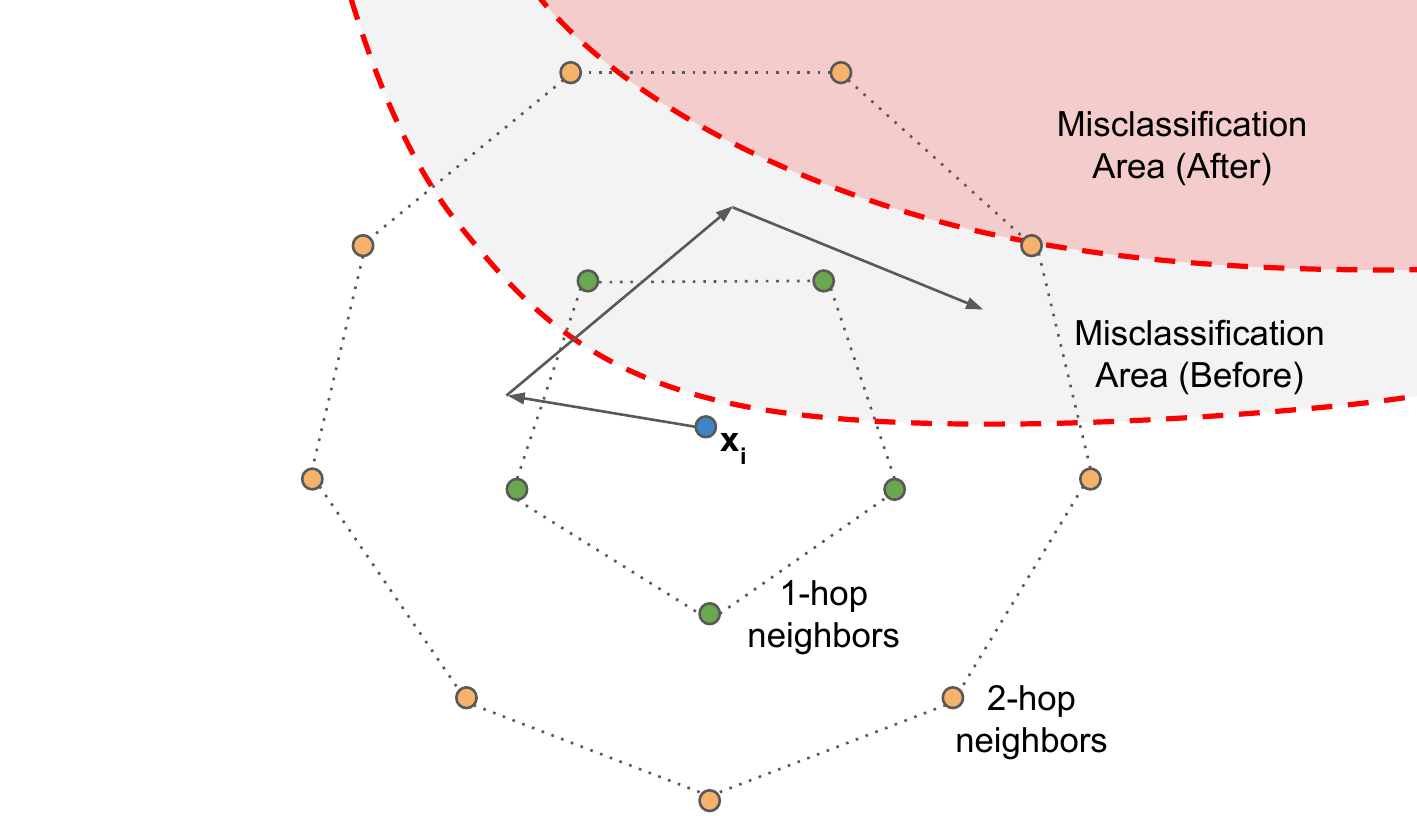}
    \caption{Dirichlet neighborhood sampling}
    \label{fig:dirichlet}
  \end{subfigure}
  \hfill
  \begin{subfigure}{0.49\textwidth}
  \centering
    \includegraphics[trim=5cm 2cm 6cm 4cm, clip, width=1.0\linewidth]{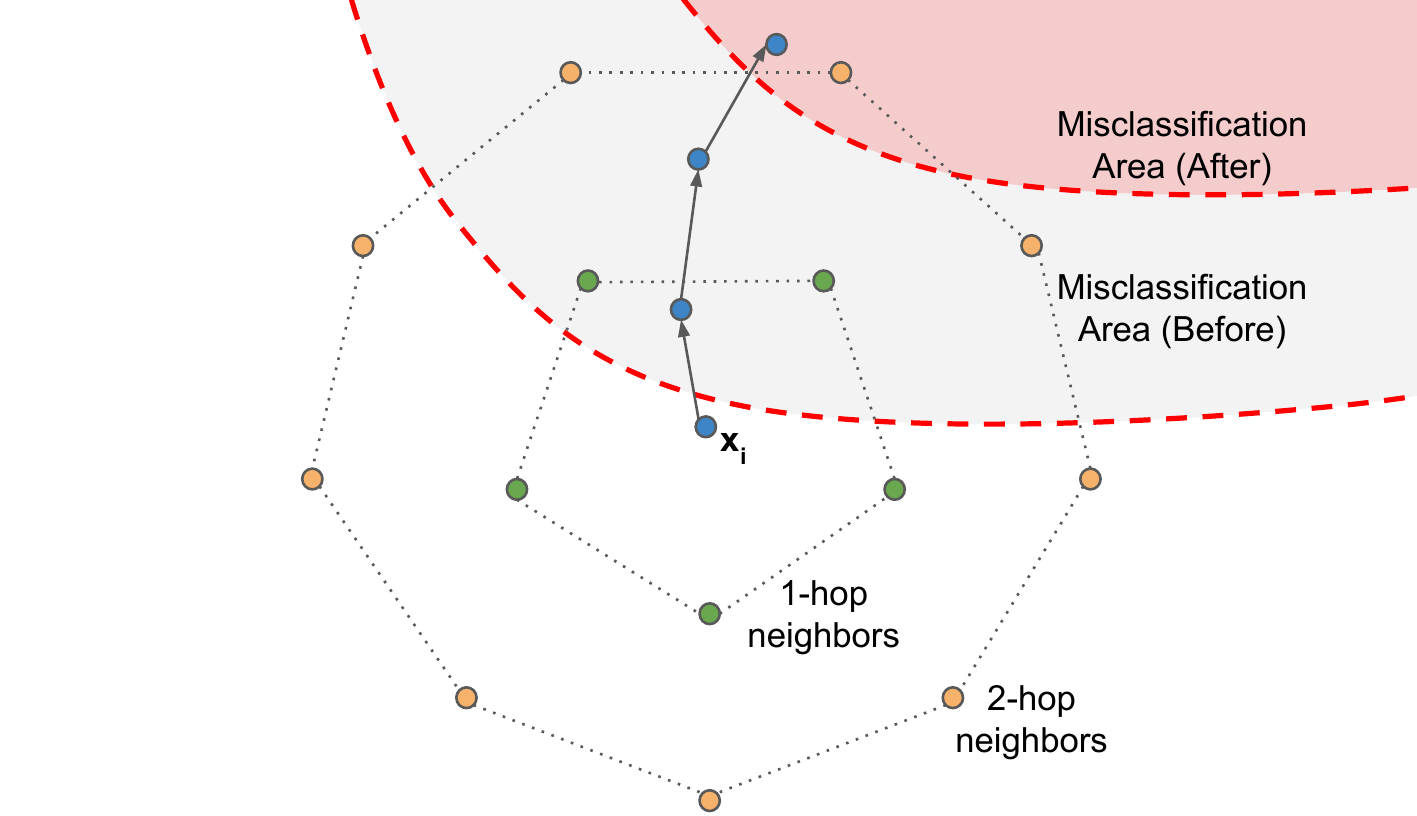}
    \caption{Adversarial Training}
    \label{fig:advtrain}
  \end{subfigure}
  \hfill
  \caption{The proposed defense methods: Dirichlet neighborhood sampling and adversarial training. Blue dots represent the target item, green dots represent 1-hop neighbors and yellow dots represent 2-hop neighbors.}
  \vspace{-10pt}
\end{figure*}

\subsection{Dirichlet Neighborhood Sampling}

To efficiently train a robust sequential recommender, we first propose Dirichlet neighborhood sampling, a randomized training approach to enhance model robustness. For neighborhood sampling, a connected graph is necessary to identify item neighbors. In NLP, synonym information can be used to compute multi-hop neighbors. However, there exists no natural neighbors or synonyms in recommender systems. Therefore, we propose to construct a neighborhood for each item in item scope $\mathcal{I}$ by computing the cosine similarity between all item pairs. We define item pair $(i, j)$ to be neighbors if item $i$ and item $j$ have a cosine similarity greater or equal to $\tau$ (i.e., $\mathrm{CosSim}(i, j) \geq \tau$). In case there exist no neighbors, items with the highest similarity values are used as 1-hop neighbors. Based on the existing 1-hop neighbors, we extend the graph to include 2-hop neighbors and construct a neighborhood dictionary to sample augmented item embeddings in training. The neighborhood dictionary is updated every $k$ epochs (we use $k$=10 in our experiments).

For each item $i$, we can compute a convex hull in the embedding space spanned by the item's multi-hop neighbors (i.e., vertices). The convex hull is used as the sampling space for the augmented item embedding. This allow us to sample from a Dirichlet distribution defined by the vertices in the convex hull to construct randomized embeddings. By sampling multiple times in the neighborhood, the recommender is less sensitive to local perturbations and learns to reject adversarial items, as illustrated in Figure~\ref{fig:dirichlet}. Formally, for item $i$ and the set of its multi-hop neighbors $\mathcal{C}_i$ ($\mathcal{C}_i = \{ c_{i,1}, c_{i,2}, \ldots, c_{i,|\mathcal{C}_i|} \}$), we represent the augmented item with $\bm{\eta}_{i}$ sampled from a Dirichlet distribution:
\begin{equation}
    \bm{\eta}_{i} = [ \eta_{i,1}, \eta_{i,2}, \ldots, \eta_{i,|\mathcal{C}_i|} ] \sim \mathrm{Dirichlet}(\alpha_1, \alpha_2, \ldots, \alpha_{|\mathcal{C}_i|}),
\end{equation}
where $\alpha$ values are hyperparameters for the Dirichlet distribution and can be empirically selected for multi-hop neighbors. Using the sampled $\bm{\eta}_{i}$, we can compute the augmented item embeddings as follows:
\begin{equation}
    \bm{f}_{e}(\bm{\eta}_{i}) = \sum_{j =1}^{|\mathcal{C}_i|} \eta_{i,j} \bm{f}_{e}(c_{i,j}).
\end{equation}

Since Dirichlet distributions are multivariate probability distributions with $\sum \bm{\eta}_{i} = 1$ and $\eta_{i, j} \geq 0, j = 1, 2, \ldots$, the sampled item is represented as a linear combination of neighbors in the convex hull, see Figure~\ref{fig:dirichlet}. We denote item $i$ with the blue dot, 1-hop neighbors with green dots and 2-hop neighbors with yellow dots. By adjusting $\alpha$ values, we can change the probability distribution in the sampling space and how far the augmented embedding can travel from item $i$. For example, when all $\alpha$ values are 0, the Dirichlet distribution only samples from nodes (i.e., neighbor items). With increasing $\alpha$ values, we can expect the sampled points approaching the center point of the convex hull.

During training, we randomly select and replace items in input sequence $\bm{x}$. Specifically, we select $p$ proportion of the items and sample $\bm{\eta}$ from a Dirichlet distribution to compute augmented embeddings using the constructed neighborhoods. The virtual sequence is then used to optimize the recommender. Given the increased number of 2-hop neighbors, we select $1.0$ as $\alpha$ for 1-hop neighbors and $0.5$ for 2-hop neighbors in our experiments~\cite{zhou2020defense}.

\subsection{Adversarial Training with Mixed Representations}
\label{sec:advtrain}

While Dirichlet neighborhood sampling is efficient for training sequential recommender, it does not exploit training examples by computing the worst case augmentation. To further improve model robustness, we design an additional defense method called adversarial training with mixed representations. Different from existing adversarial training methods~\cite{tang2019adversarial, zhu2019freelb} and Dirichlet neighborhood sampling, the proposed method searches for the worst case linear combination of items $\bm{\eta}$ in $\mathcal{I}$ instead of sampling from the local neighborhood alone. The idea behind searching for item combinations is to introduce mixed representations of the original items and potential adversarial items. Then, we can compute the worst case linear combinations by updating $\bm{\eta}$ using gradient ascent, such that both the original and adversarial item embeddings are jointly optimized for robust item representations.

For item $x_i$ in input $\bm{x}$, we first initialize $x_i$ with one-hot vector $\bm{\eta}_{i} = \mathrm{Onehot}(x_i) \in \mathcal{R}^{|\mathcal{I}|}$. Similar to Dirichlet neighborhood sampling, $\bm{f}_{e}(\bm{\eta}_{i})$ is computed as the weighted sum of the embedding matrix (i.e., $\bm{f}_{e}(\bm{\eta}_{i}) = \sum_{j=1}^{|\mathcal{I}|} \bm{\eta}_{i,j} \bm{f}_{e}(j)$).
Next, we perform gradient ascent and compute the gradients w.r.t. $\bm{\eta}_{i}$. Specifically, the embedded sequence $\tilde{\bm{x}}$ for input $\bm{x}$ is first computed using the embedding function $\bm{f}_e$. Similar to Section~\ref{sec:attack}, we then forward $\tilde{\bm{x}}$ using the sequential model $\bm{f}_m$ and compute the gradients based on the optimization loss to update $\bm{\eta}_{i}$:
\begin{equation}
    \bm{\eta}_{i} = \bm{\eta}_{i} + \epsilon \frac{\nabla_{\bm{\eta}_{i}}\mathcal{L}(\bm{f}_{m}(\tilde{\bm{x}}), y)}{\| \nabla_{\bm{\eta}_{i}} \mathcal{L}(\bm{f}_{m}(\tilde{\bm{x}}), y) \|},
\end{equation}
where $\epsilon$ is the step size. In each update, we normalize the gradients by dividing the gradients by their $l^2$ norm. Then, step size $\epsilon$ is multiplied with the gradients before performing gradient ascent. As such, $\bm{\eta}_{i}$ is perturbed towards the misclassification area of the sequential recommender, see Figure~\ref{fig:advtrain}. As $\bm{\eta}_{i}$ is a linear combination of all items in $\mathcal{I}$, we require $\sum_{j}^{|\mathcal{I}|} \bm{\eta}_{i, j} = 1$ and $\bm{\eta}_{i, j} \geq 0, j = 1, 2, \ldots$ and ensure that the updated $\bm{\eta}_{i}$ follows the same requirements within the embedding space. Therefore, we clip negative elements and rescale $\bm{\eta}_{i}$ after each update\footnote{We considered using softmax function but eventually discarded the idea as softmax only produces non-zero $\eta$ values.}:
\begin{equation}
    \bm{\eta}_{i} = \frac{\mathrm{Clip}_{\left[ 0, \infty \right]}(\bm{\eta}_{i}, 0, 1)}{\mathrm{Sum}(\mathrm{Clip}_{\left[ 0, \infty \right]}(\bm{\eta}_{i}, 0, 1))}.
\end{equation}


We illustrate the proposed adversarial training with mixed representations in Figure~\ref{fig:advtrain}. We denote item $i$ with the blue dot, the paths of the blue dot represent the update iterations of adversarial training, in which the adversarial example moves towards the misclassification area, and thereby forcing the recommender to resist against worst case perturbations. Unlike Dirichlet sampling, where the augmentation directions are sampled from a distribution within a local convex hull, adversarial training finds the worst case embeddings using a linear combination of items to propagate gradients to the item representations. Such worst case augmentation extends the model robustness to a larger area outside the local neighborhood, and therefore effectively reduces the area of misclassification. Compared to Dirichlet neighborhood sampling, adversarial training with mixed representations requires increased computational costs, since adversarial examples are updated multiple times in each iteration. To reduce the training costs, we select $p$ proportion of items in each sequence and update $\bm{\eta}$ of the selected items three times in our experiments, we optimize the recommender with both clean input and perturbed examples to enhance the model robustness.
\section{Experiments}

\begin{table}[t]
\parbox{.49\linewidth}{
\small
\centering
\begin{tabular}{cccccc}
\toprule
\textbf{Datasets} & \textbf{Users} & \textbf{Items} & \textbf{Avg. len} & \textbf{Sparsity} \\ \midrule
\textbf{Beauty}   & 22,363         & 12,101         & 9                 & 99.9\%            \\
\textbf{ML-1M}    & 6,040          & 3,416          & 165               & 95.2\%            \\
\textbf{ML-20M}   & 138,493        & 18,345         & 144               & 99.2\%            \\
\textbf{LastFM}   & 988            & 57,638         & 9739              & 83.1\%            \\
\textbf{Steam}    & 334,289        & 12,012         & 13                & 99.9\%            \\
 \bottomrule
\end{tabular}
\caption{Dataset details.}
\label{tab:datasets}
}
\hfill
\parbox{.49\linewidth}{
\small
\centering
\begin{tabular}{ccc}
\toprule
\textbf{Model}                           & \textbf{Basic Block} & \textbf{Training} \\ \midrule
\textbf{NARM~\cite{li2017neural}}        & RNN                  & Autoregressive    \\
\textbf{SASRec~\cite{kang2018self}}      & Transformer          & Autoregressive    \\
\textbf{BERT4Rec~\cite{sun2019bert4rec}} & Transformer          & Masked Training   \\
\textbf{Locker~\cite{he2021locker}}      & Transformer          & Masked Training   \\
\bottomrule
\end{tabular}
\caption{Model Architectures.}
\label{tab:models}
}
\vspace{-10pt}
\end{table}

\subsection{Setup}
\subsubsection{Dataset} In our experiments, we adopt five datasets to validate the proposed attack and defense methods (Details presented in Table~\ref{tab:datasets}). We use Amazon Beauty~\cite{ni-etal-2019-justifying}, Movielens (both Movielens 1M and Movielens 20M)~\cite{harper2015movielens}, LastFM~\cite{Celma:Springer2010} and Steam~\cite{mcauley2015image}. We adopt 5-core version of all datasets and preprocess the data following~\cite{sun2019bert4rec, yue2021black}. 

\subsubsection{Model} We select four different recommender architectures to evaluate the proposed methods, see Table~\ref{tab:models}. In particular, we adopt NARM~\cite{li2017neural}, SASRec~\cite{kang2018self}, BERT4Rec~\cite{sun2019bert4rec} and Locker~\cite{he2021locker} in our experiments: 
\begin{itemize}
    \item \emph{Neural Attentive Recommendation Machine (NARM)} is a RNN-based sequential recommender comprising of a global encoder and a local encoder. NARM utilizes an attention module to compute item representations and outputs the predictions using a similarity layer~\cite{li2017neural}.
    \item \emph{Self-Attentive Sequential Recommendation (SASRec)} leverages transformer blocks to forward items autoregressively. Transformer blocks compute attentive representations with one-directional self-attention. SASRec consists of an embedding function, transformer blocks and an output layer~\cite{kang2018self}.
    \item \emph{Bidirectional Encoder Representations from Transformers for Sequential Recommendation (BERT4Rec)} has an similar architecture to SASRec. BERT4Rec utilizes bidirectional self-attention and adopts masked training (i.e., masked language modeling) to optimize the model~\cite{sun2019bert4rec}.
    \item \emph{Locally Constrained Self-attentive Recommender (Locker)} is similar to BERT4Rec, but proposes additional local constraints to improve self-attention. In our implementation, we use the convolution layers to model local dynamics. Locker is also optimized via masked training~\cite{he2021locker}.
\end{itemize}

\subsubsection{Evaluation} We follow~\cite{kang2018self, sun2019bert4rec, yue2021black} to perform evaluation. We adopt leave-one-out method and use the last two items in each sequence for validation and testing. We adopt normalized discounted cumulative gain \emph{NDCG@10} and \emph{Recall@10} as evaluation metrics. 
In the attack experiments, 15 items are selected as target items across different popularity groups for targeted attacks, we evaluate on the polluted profiles and compute the average results.

\subsubsection{Implementation} We follow the original papers to implement sequential recommenders~\cite{li2017neural, kang2018self, sun2019bert4rec, he2021locker}. Unless explicitly mentioned, hyperparameters are from the original works. All models are trained without warmup using Adam optimizer with learning rate of 0.001, weight decay 0.01 and batch size of 64. Similar to~\cite{kang2018self, sun2019bert4rec, yue2021black}, we set the maximum sequence lengths of ML-1M to be 200 and 50 for other datasets. For profile pollution, we use $z=2$ as the maximum substitution for ML-1M and and $z = 1$ for other datasets. The minimum cosine similarity of $\tau = 0.5$ is imposed for both attack and defense methods.
In our defense experiments, we set substitution probability $p$ to be $0.5$ and adopt 2-hop neighborhood for Dirichlet sampling. To update perturbations in adversarial training, an update step size $\epsilon$ from $\left[0.01, 0.1, 1.0 \right]$ and three adversarial update iterations are used in our implementation\footnote{The implementation of our framework is publicly available at https://github.com/Yueeeeeeee/RecSys-Substitution-Defense.}. 

\subsection{RQ1: Can We Attack Recommenders using Substitution-based Profile Pollution?}

\begin{table}[t]
\begin{tabular}{@{}lcccccc@{}}
\toprule
\multirow{2}{*}{\textbf{Model (U/T)}} & \textbf{Dataset} & \textbf{Beauty}      & \textbf{ML-1M}       & \textbf{ML-20M}      & \textbf{LastFM}      & \textbf{Steam}       \\ \cmidrule(l){2-7} 
                                      & Method           & N@10 / R@10            & N@10 / R@10            & N@10 / R@10            & N@10 / R@10            & N@10 / R@10            \\ \midrule
\multirow{3}{*}{NARM (U)}             & Before           & 0.334 / 0.493          & 0.624 / 0.815          & 0.770 / 0.940          & 0.659 / 0.804          & 0.617 / 0.836          \\
                                      & SimAlter         & 0.326 / 0.483          & 0.622 / 0.813          & 0.769 / 0.939          & 0.654 / 0.795          & 0.611 / 0.831          \\
                                      & Ours             & \textbf{0.209 / 0.332} & \textbf{0.154 / 0.326} & \textbf{0.630 / 0.892} & \textbf{0.565 / 0.727} & \textbf{0.437 / 0.714} \\
\multirow{3}{*}{NARM (T)}             & Before           & 0.064 / 0.125          & 0.074 / 0.150          & 0.093 / 0.157          & 0.099 / 0.147          & 0.088 / 0.141          \\
                                      & SimAlter         & 0.094 / 0.170          & 0.084 / 0.169          & 0.098 / 0.165          & 0.100 / 0.148          & 0.101 / 0.158          \\
                                      & Ours             & \textbf{0.239 / 0.420} & \textbf{0.285 / 0.436} & \textbf{0.140 / 0.223} & \textbf{0.262 / 0.407} & \textbf{0.181 / 0.293} \\ \midrule
\multirow{3}{*}{SASRec (U)}           & Before           & 0.325 / 0.498          & 0.546 / 0.790          & 0.762 / 0.953          & 0.679 / 0.823          & 0.599 / 0.820          \\
                                      & SimAlter         & 0.320 / 0.491          & 0.545 / 0.788          & 0.760 / 0.952          & 0.680 / 0.824          & 0.591 / 0.813          \\
                                      & Ours             & \textbf{0.185 / 0.309} & \textbf{0.144 / 0.299} & \textbf{0.640 / 0.900} & \textbf{0.589 / 0.771} & \textbf{0.407 / 0.641} \\
\multirow{3}{*}{SASRec (T)}           & Before           & 0.072 / 0.139          & 0.092 / 0.165          & 0.104 / 0.173          & 0.103 / 0.152          & 0.093 / 0.149          \\
                                      & SimAlter         & 0.184 / 0.312          & 0.109 / 0.192          & 0.124 / 0.211          & 0.145 / 0.231          & 0.161 / 0.281          \\
                                      & Ours             & \textbf{0.187 / 0.335} & \textbf{0.236 / 0.360} & \textbf{0.139 / 0.226} & \textbf{0.191 / 0.296} & \textbf{0.179 / 0.297} \\ \midrule
\multirow{3}{*}{BERT4Rec (U)}         & Before           & 0.353 / 0.530          & 0.542 / 0.744          & 0.760 / 0.941          & 0.663 / 0.796          & 0.574 / 0.795          \\
                                      & SimAlter         & 0.340 / 0.516          & 0.540 / 0.742          & 0.759 / 0.940          & 0.660 / 0.796          & 0.563 / 0.788          \\
                                      & Ours             & \textbf{0.215 / 0.349} & \textbf{0.155 / 0.316} & \textbf{0.636 / 0.892} & \textbf{0.637 / 0.778} & \textbf{0.305 / 0.566} \\
\multirow{3}{*}{BERT4Rec (T)}         & Before           & 0.072 / 0.147          & 0.064 / 0.133          & 0.086 / 0.155          & 0.106 / 0.157          & 0.087 / 0.139          \\
                                      & SimAlter         & 0.084 / 0.170          & 0.072 / 0.149          & 0.088 / 0.156          & 0.106 / 0.158          & 0.093 / 0.150          \\
                                      & Ours             & \textbf{0.155 / 0.298} & \textbf{0.245 / 0.395} & \textbf{0.114 / 0.184} & \textbf{0.119 / 0.181} & \textbf{0.144 / 0.247} \\ \midrule
\multirow{3}{*}{Locker (U)}           & Before           & 0.357 / 0.526          & 0.602 / 0.796          & 0.760 / 0.941          & 0.662 / 0.799          & 0.631 / 0.852          \\
                                      & SimAlter         & 0.353 / 0.520          & 0.601 / 0.796          & 0.759 / 0.941          & 0.660 / 0.804          & 0.627 / 0.852          \\
                                      & Ours             & \textbf{0.248 / 0.389} & \textbf{0.164 / 0.346} & \textbf{0.653 / 0.902} & \textbf{0.617 / 0.768} & \textbf{0.507 / 0.780} \\
\multirow{3}{*}{Locker (T)}           & Before           & 0.065 / 0.130          & 0.073 / 0.147          & 0.084 / 0.150          & 0.104 / 0.157          & 0.097 / 0.159          \\
                                      & SimAlter         & 0.133 / 0.242          & 0.084 / 0.167          & 0.090 / 0.159          & 0.109 / 0.166          & \textbf{0.135 / 0.230} \\
                                      & Ours             & \textbf{0.134 / 0.250} & \textbf{0.234 / 0.356} & \textbf{0.107 / 0.180} & \textbf{0.126 / 0.200} & 0.131 / 0.225          \\ \bottomrule
\end{tabular}
\caption{Evaluation results of the clean performance and performance under profile pollution attacks. Untargeted attack results are denoted with U (lower the better), targeted attacks are denoted with T (higher the better). Before denotes clean performance, while SimAlter and Ours represent the model performance under attack.}
\label{tab:attack-results}
\vspace{-10pt}
\end{table}

We first evaluate the attack performance of the proposed profile pollution algorithm on all model architectures and datasets, the results are presented in Table~\ref{tab:attack-results}. Since we restrain from substituting more than two items, we adopt SimAlter as the baseline method, which utilizes neighbor items as adversarial items to perform attacks~\cite{yue2021black}. The table includes several parts: (1)~Each row denotes a model name and the setting (i.e., U for untargeted and T for targeted attacks, Before denotes the clean performance); (2)~Each column represents the results from one dataset; (3)~For targeted attacks, we select a total of 15 items across different popularity groups and report the average results; (4)~\emph{Lower results indicate better untargeted attack performance while higher results indicate better targeted attack performance}; and (5)~The presented results are in \emph{NDCG@10} and \emph{Recall@10} (i.e., N@10/R@10). Best results are marked in bold.

From the results we observe the following: (1)~By comparing clean performance (i.e., Before) and metric numbers under attack (i.e., SimAlter and Ours), recommender models demonstrate vulnerability against profile pollution attacks. For example, recommenders suffer from over $50\%$ performance deterioration in untargeted attacks on ML-1M. (2)~The proposed method performs the best for both untargeted and targeted attacks, successfully biases the recommender models towards the targets and outperforms the baseline method in all cases with the only exception of Locker on Steam. (3)~We observe different vulnerabilities across model architectures and training datasets. For example, we observe RNN-based model (i.e., NARM) to be more vulnerable against targeted attacks with an average increase of over $175\%$ on NDCG@10. On the contrary, Locker only suffers from $82.1\%$ increase in targeted attacks. Additionally, we observe that dense datasets are generally more robust against profile pollution attacks. For instance, we only observe $9.5\%$ performance drop for untargeted attacks on LastFM dataset, while the same setting yields $37.5\%$ deterioration on Beauty. Overall, the results suggest that the proposed algorithm can effectively perform profile pollution attacks.

\subsection{RQ2: How the Proposed Defense Methods Perform under Profile Pollution Attacks?}

\begin{table}[t]
\begin{tabular}{@{}lcccccc@{}}
\toprule
\multirow{2}{*}{\textbf{Model (U/T)}} & \textbf{\textbf{Dataset}} & \textbf{\textbf{Beauty}} & \textbf{\textbf{ML-1M}}  & \textbf{\textbf{ML-20M}} & \textbf{\textbf{LastFM}} & \textbf{\textbf{Steam}}  \\ \cmidrule(l){2-7} 
                                      & Method                    & $\Delta$ N@10/R@10       & $\Delta$ N@10/R@10       & $\Delta$ N@10/R@10       & $\Delta$ N@10/R@10       & $\Delta$ N@10/R@10       \\ \midrule
\multirow{3}{*}{NARM (U)}             & Before                    & -0.125 / -0.161          & -0.470 / -0.489          & -0.140 / -0.048          & -0.094 / -0.077          & -0.180 / -0.122          \\
                                      & Dirichlet                 & -0.100 / -0.120          & -0.366 / -0.335          & -0.136 / -0.060          & \textbf{-0.031 / -0.018} & -0.160 / -0.098          \\
                                      & AdvTrain                  & \textbf{-0.093 / -0.120} & \textbf{-0.331 / -0.314} & \textbf{-0.100 / -0.035} & -0.053 / -0.043          & \textbf{-0.159 / -0.102} \\
\multirow{3}{*}{NARM (T)}             & Before                    & 0.175 / 0.295            & 0.211 / 0.286            & 0.047 / 0.066            & 0.163 / 0.260            & 0.093 / 0.152            \\
                                      & Dirichlet                 & 0.127 / 0.215            & 0.203 / 0.270            & 0.059 / 0.079            & \textbf{0.040 / 0.078}   & 0.071 / 0.114            \\
                                      & AdvTrain                  & \textbf{0.092 / 0.156}   & \textbf{0.153 / 0.201}   & \textbf{0.020 / 0.024}   & 0.050 / 0.080            & \textbf{0.062 / 0.093}   \\ \midrule
\multirow{3}{*}{SASRec (U)}           & Before                    & -0.140 / -0.189          & -0.402 / -0.491          & -0.122 / -0.053          & -0.090 / -0.052          & -0.192 / -0.179          \\
                                      & Dirichlet                 & -0.127 / -0.180          & -0.299 / -0.307          & -0.091 / -0.031          & -0.038 / -0.030          & -0.122 / -0.096          \\
                                      & AdvTrain                  & \textbf{-0.093 / -0.121} & \textbf{-0.228 / -0.215} & \textbf{-0.067 / -0.021} & \textbf{-0.031 / -0.017} & \textbf{-0.120 / -0.085} \\
\multirow{3}{*}{SASRec (T)}           & Before                    & 0.115 / 0.196            & 0.144 / 0.195            & 0.035 / 0.053            & 0.088 / 0.144            & 0.086 / 0.148            \\
                                      & Dirichlet                 & 0.089 / 0.148            & 0.097 / 0.122            & 0.023 / 0.032            & \textbf{0.027 / 0.051}   & \textbf{0.044 / 0.087}   \\
                                      & AdvTrain                  & \textbf{0.061 / 0.096}   & \textbf{0.080 / 0.109}   & \textbf{0.021 / 0.035}   & 0.041 / 0.082            & 0.050 / 0.093            \\ \midrule
\multirow{3}{*}{BERT4Rec (U)}         & Before                    & -0.138 / -0.181          & -0.387 / -0.428          & -0.124 / -0.049          & -0.026 / -0.018          & -0.269 / -0.229          \\
                                      & Dirichlet                 & \textbf{-0.071 / -0.095} & -0.258 / -0.228          & -0.081 / -0.031          & -0.011 / -0.009          & \textbf{-0.148 / -0.094} \\
                                      & AdvTrain                  & -0.119 / -0.152          & \textbf{-0.249 / -0.234} & \textbf{-0.074 / -0.029} & \textbf{-0.006 / -0.008} & -0.151 / -0.106          \\
\multirow{3}{*}{BERT4Rec (T)}         & Before                    & 0.083 / 0.151            & 0.181 / 0.262            & 0.028 / 0.029            & 0.013 / 0.024            & 0.057 / 0.108            \\
                                      & Dirichlet                 & 0.041 / 0.069            & 0.129 / 0.162            & 0.021 / 0.027            & 0.005 / 0.010            & \textbf{0.026 / 0.049}   \\
                                      & AdvTrain                  & \textbf{0.056 / 0.096}   & \textbf{0.132 / 0.172}   & \textbf{0.020 / 0.019}   & \textbf{0.001 / 0.009}   & 0.035 / 0.068            \\ \midrule
\multirow{3}{*}{Locker (U)}           & Before                    & -0.109 / -0.137          & -0.438 / -0.450          & -0.107 / -0.039          & -0.045 / -0.031          & -0.124 / -0.072          \\
                                      & Dirichlet                 & -0.087 / -0.106          & \textbf{-0.301 / -0.262} & -0.093 / -0.034          & -0.040 / -0.027          & -0.125 / -0.072          \\
                                      & AdvTrain                  & \textbf{-0.086 / -0.106} & -0.314 / -0.279          & \textbf{-0.084 / -0.031} & \textbf{-0.028 / -0.019} & \textbf{-0.117 / -0.066} \\
\multirow{3}{*}{Locker (T)}           & Before                    & 0.069 / 0.120            & 0.161 / 0.209            & 0.023 / 0.030            & 0.022 / 0.043            & 0.034 / 0.066            \\
                                      & Dirichlet                 & 0.066 / 0.100            & \textbf{0.121 / 0.157}   & \textbf{0.021 / 0.026}   & 0.019 / 0.037            & 0.021 / 0.043            \\
                                      & AdvTrain                  & \textbf{0.044 / 0.071}   & 0.131 / 0.166            & 0.024 / 0.027            & \textbf{0.012 / 0.023}   & \textbf{0.015 / 0.022}   \\ \bottomrule
\end{tabular}
\caption{Evaluation results of the defense methods under profile pollution attacks. Untargeted attack results are denoted with U (higher the better), targeted attacks are denoted with T (lower the better). Before denotes evaluation results on the clean model, while Dirichlet and AdvTrain represent the adopted defense methods.}
\label{tab:defense-results}
\vspace{-10pt}
\end{table}

We study the effectiveness of defense methods by evaluating profile pollution attacks on recommenders trained with the defense methods. In particular, we adopt identical training conditions and apply the proposed methods (i.e., Dirichlet neighborhood sampling and adversarial training). Then, we evaluate the recommender performance on clean test data and polluted test data. As each defense method results in a new model, we only compute the \textbf{performance changes ($\Delta$)} between clean and polluted evaluation data on the model. Ideally, a robust recommender demonstrates little performance variations, therefore lower absolute values of $\Delta$ indicate better defense performance. In other words, higher values indicate better results for untargeted attacks, as model degrades after attack (i.e., $\Delta$ is negative). Lower values indicate better results for targeted attacks, as metric values w.r.t. target items increase after attack (i.e., $\Delta$ is positive), see Table~\ref{tab:defense-results}. Similar to Table~\ref{tab:attack-results}, we separate different settings by model, dataset and attack type. We adopt $\Delta$ of \emph{NDCG@10} and \emph{Recall@10} as metrics and mark best results in bold.

The defense results indicate the following: (1)~Although the recommenders are vulnerable against profile pollution attacks, we find both defense methods to be helpful against polluted test data. For example, the proposed adversarial training reduces NDCG@10 deterioration in untargeted attacks by $34.2\%$ on ML-1M. (2)~The results indicate that the proposed adversarial training performs the best, achieving the best performance in 30 out of 40 scenarios. Overall, adversarial training reduces NDCG@10 variations by $34.4\%$ in untargeted attacks, compared to $28.5\%$ of the proposed Dirichlet neighborhood sampling method. The results suggest adversarial training considerably improves model robustness and can outperform randomized method (i.e., Dirichlet sampling) in most cases. This may be attributed to the generated adversarial examples, which forces the model to reduce reliance on vulnerable features. (3)~Defense methods perform differently depending on the recommender and datasets. For example, the proposed adversarial training demonstrate better performance on NARM and SASRec, consistently outperforming Dirichlet neighborhood sampling on four of the five datasets. A potential reason may be the autoregressive training approach, which increases model vulnerability by learning shortcuts and local features~\cite{geirhos2020shortcut}. (4)~Based on the defense results, LastFM demonstrates the highest robustness on all datasets against both types of attacks, while Locker performs the best across all recommender architectures. This suggest that increasing data density and capturing user dynamics contribute to robustness, possibly by providing more informative item representations and transition patterns. Moreover, learning via autoencoding (i.e., masked training) might improve recommender robustness against adversarial attacks.

\begin{table}[ht]
\begin{minipage}[b]{0.59\linewidth}
\centering
\small
\begin{tabular}{@{}lccc@{}}
\toprule
\textbf{Popularity} & \textbf{Popular}     & \textbf{Middle}      & \textbf{Bottom}      \\ \cmidrule(l){2-4} 
Dataset             & $\Delta$ N@10 / R@10 & $\Delta$ N@10 / R@10 & $\Delta$ N@10 / R@10 \\ \midrule
Beauty Before       & 0.178 / 0.252        & 0.087 / 0.177        & 0.067 / 0.143        \\
Beauty AdvTrain     & 0.126 / 0.177        & 0.041 / 0.075        & 0.021 / 0.049        \\ \midrule
ML-1M Before        & 0.409 / 0.478        & 0.057 / 0.118        & 0.057 / 0.119        \\
ML-1M AdvTrain      & 0.318 / 0.368        & 0.037 / 0.081        & 0.022 / 0.051        \\ \midrule
ML-20M Before       & 0.081 / 0.089        & 0.015 / 0.036        & 0.004 / 0.009        \\
ML-20M AdvTrain     & 0.066 / 0.078        & 0.007 / 0.017        & 0.001 / 0.003        \\ \midrule
LastFM Before       & 0.105 / 0.152        & 0.058 / 0.107        & 0.053 / 0.094        \\
LastFM AdvTrain     & 0.058 / 0.086        & 0.015 / 0.034        & 0.010 / 0.025        \\ \midrule
Steam Before        & 0.138 / 0.226        & 0.041 / 0.084        & 0.023 / 0.047        \\
Steam AdvTrain      & 0.110 / 0.186        & 0.010 / 0.029        & 0.004 / 0.009        \\ \bottomrule
\end{tabular}
\captionof{table}{Averaged results of different popularity groups in targeted attacks.}
\label{tab:popularity_dataset}
\end{minipage}\hfill
\begin{minipage}[b]{0.39\linewidth}
\centering
\includegraphics[width=\linewidth]{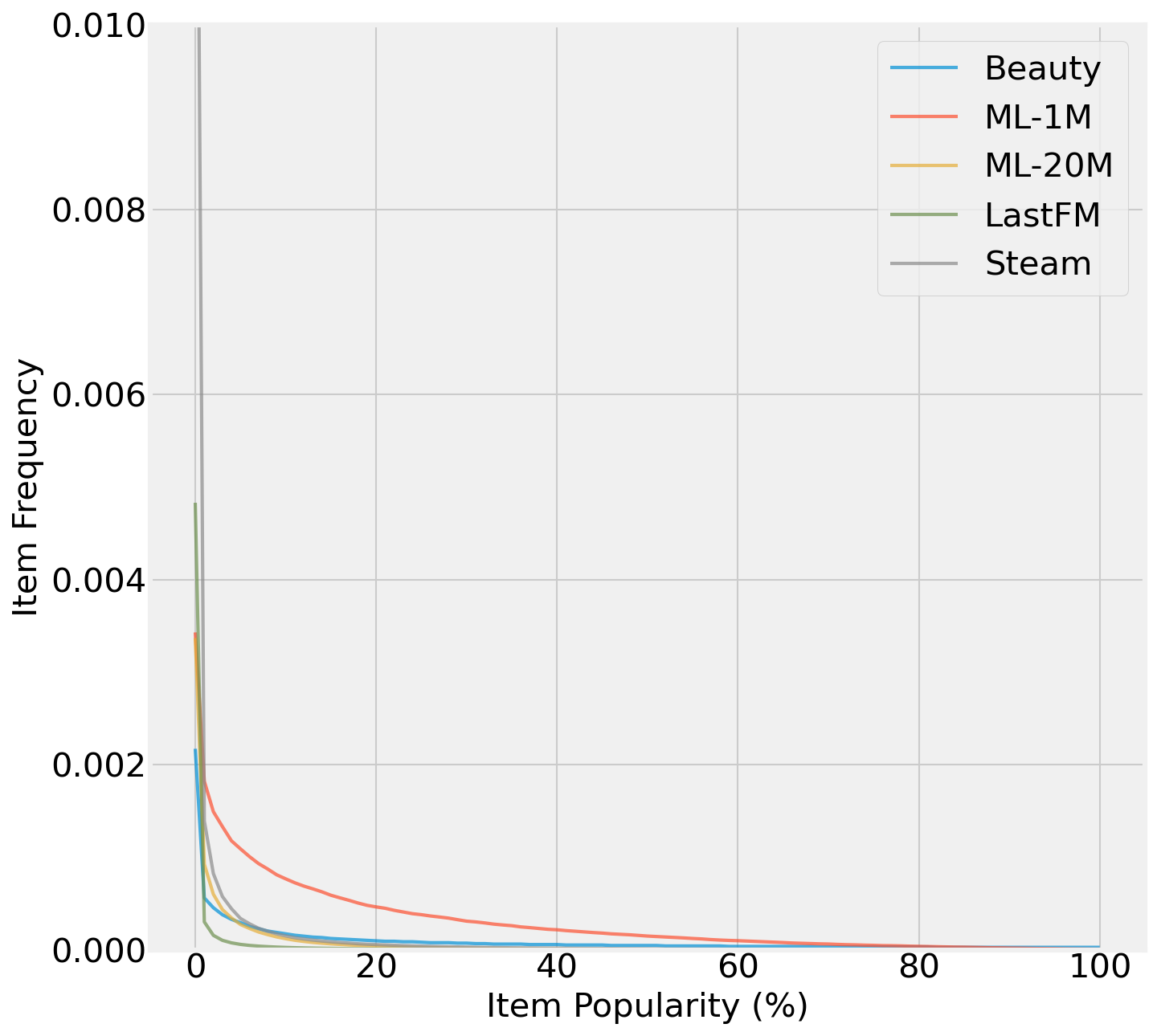}  
\captionof{figure}{Distribution of Item Popularity.}
\label{fig:distribution}
\end{minipage}
\vspace{-10pt}
\end{table}

We additionally study the performance variation w.r.t. item popularity. Specifically, we study targeted attacks on sequential recommenders and select target items of different popularity. Target items are divided into three subgroups upon their popularity (i.e., number of total occurrences) in the dataset. \emph{Popular} items denote the $\frac{1}{3}$ most popular items within the item scope. \emph{Middle} and \emph{Bottom} items refer to the next $\frac{1}{3}$ and the last $\frac{1}{3}$ of the items in $\mathcal{I}$. We present the item distribution of all datasets in Figure~\ref{fig:distribution}, we present profile pollution results of the three popularity subgroups averaged across all recommender models. Similarly, performance variations are report with $\Delta$ of \emph{NDCG@10} and \emph{Recall@10} in Table~\ref{tab:popularity_dataset}.  We observe the following: (1)~Popular items are of the highest vulnerability with $\Delta$ N@10 of 0.182 before attack, showing the largest performance variations in all cases; (2)~Middle and bottom items are generally `harder' to attack. For example, middle items has a much lower $\Delta$ N@10 of 0.051 before attack. (3)~Popular items are also `harder' to defend, with $27.0\%$ reduced $\Delta$ N@10 after defense, compared to $58.2\%$ and $73.8\%$ of middle and bottom items. The results suggest that popular items exhibit the highest vulnerability against profile pollution attacks, a potential reason for such vulnerability is the higher number of neighbors, which increases the number of possible adversarial substitutes.

\subsection{RQ3: How Recommender Performance and Robustness Change with Different Model Architectures?}

\begin{table}[t]
\centering
\begin{tabular}{@{}lccccc@{}}
\toprule
\textbf{Popularity} & \textbf{Performance} & \textbf{Untargeted}  & \textbf{Popular}     & \textbf{Middle}      & \textbf{Bottom}      \\ \midrule
Model               & N@10 / R@10          & $\Delta$ N@10 / R@10 & $\Delta$ N@10 / R@10 & $\Delta$ N@10 / R@10 & $\Delta$ N@10 / R@10 \\ \midrule
NARM Before         & 0.601/0.778          & -0.202 / -0.179      & 0.241 / 0.303        & 0.099 / 0.188        & 0.073 / 0.144        \\
NARM AdvTrain       & 0.576/0.766          & -0.148 / -0.124      & 0.187 / 0.238        & 0.032 / 0.076        & 0.011 / 0.027        \\ \midrule
SASRec Before       & 0.582/0.777          & -0.189 / -0.193      & 0.182 / 0.239        & 0.063 / 0.127        & 0.037 / 0.095        \\
SASRec AdvTrain     & 0.585/0.779          & -0.108 / -0.092      & 0.110 / 0.157        & 0.027 / 0.055        & 0.018 / 0.040        \\ \midrule
BERT4Rec Before     & 0.578/0.761          & -0.189 / -0.181      & 0.157 / 0.218        & 0.025 / 0.054        & 0.036 / 0.072        \\
BERT4Rec AdvTrain   & 0.579/0.771          & -0.121 / -0.105      & 0.127 / 0.172        & 0.015 / 0.023        & 0.010 / 0.022        \\ \midrule
Locker Before       & 0.602/0.783          & -0.165 / -0.146      & 0.149 / 0.196        & 0.020 / 0.048        & 0.016 / 0.037        \\
Locker AdvTrain     & 0.593/0.778          & -0.129 / -0.102      & 0.117 / 0.148        & 0.015 / 0.035        & 0.007 / 0.018        \\ \bottomrule
\end{tabular}
\captionof{table}{Influence of recommender architecture on profile pollution attacks.}
\label{tab:popularity_model}
\vspace{-10pt}
\end{table}

\begin{figure}[t]
  \begin{subfigure}{\textwidth}
    \centering
    \includegraphics[width=\linewidth]{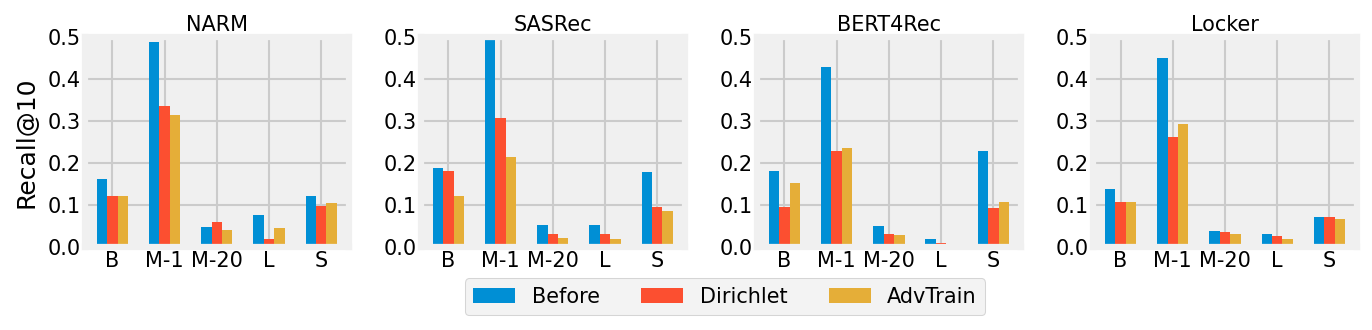}
    \caption{Average results of untargeted attacks with different defense methods.}
    \label{fig:untargeted}
  \end{subfigure}
  \begin{subfigure}{\textwidth}
    \centering
    \includegraphics[width=\linewidth]{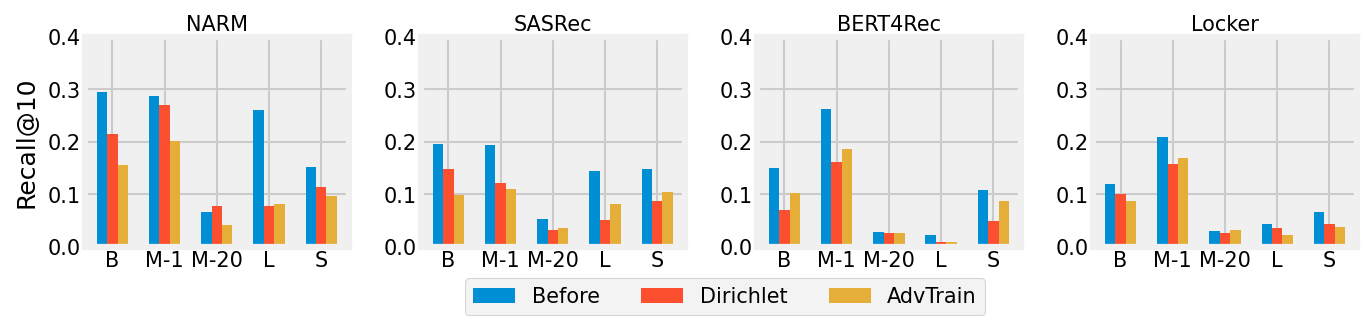}
    \caption{Average results of targeted attacks with different defense methods.}
    \label{fig:targeted}
  \end{subfigure}
  \caption{Visualization of profile pollution performance on different recommenders / datasets. B, M-1, M-20, L, S represent Beauty, ML-1M, ML-20M, LastFM and Steam respectively.}
  \vspace{-10pt}
\end{figure}

We study the vulnerability of different recommender models and performance of the proposed defense methods. In particular, we study the attacks on all datasets and compare the original performance and performance changes of different recommender architectures under profile pollution attacks. We present the averaged results of each recommender architecture in Table~\ref{tab:popularity_model} for a direct comparison. The visualized results are presented in Figure~\ref{fig:untargeted} and Figure~\ref{fig:targeted}. Each sub-graph represents a recommender model, where the results on all datasets with different defense methods are compared and visualized with bars. As NDCG@10 and Recall@10 provide similar trends, we visualize the recall results in the graphs. B, M-1, M-20, L, S represent Beauty, ML-1M, ML-20M, LastFM and Steam datasets.

Here are our observations: (1)~We observe that adversarial training has limited influence on recommendation performance. For example, our results on transformer-based models demonstrate slight performance changes, possibly due to the improved item representations. (2)~By comparing performance variations before and after profile pollution attacks in Table~\ref{tab:popularity_model}, we observe that NARM is more vulnerable and demonstrates higher performance variations both before and after applying the defense methods. (3)~Locker demonstrates the highest robustness on average. Without adopting defense methods, we observe $27.4\%$ NDCG@10 variation on Locker in untargeted attacks, compared to $32.7\%$, $32.4\%$ and $33.6\%$ of BERT4Rec, SASRec and NARM. (4)~Although SASRec and BERT4Rec have similar transformer blocks to Locker, SASRec and BERT4Rec are more vulnerable and suffer higher performance deterioration even when adversarially trained. For instance, targeted attacks on popular items increase NDCG@10 variation by $31.3\%$ on SASRec and $27.2\%$ on BERT4Rec, compared to only $24.8\%$ of Locker. Overall, transformer-based sequential recommenders (i.e., SASRec, BERT4Rec and Locker) demonstrate improved robustness of similar magnitude compared to RNN-based model (i.e., NARM). Comparable trends can be observed when we apply adversarial training on sequential recommenders. The reason for such robustness may be traced back to the introduction of global attention in transformer models. Additionally, BERT4Rec and Locker utilize masked training to improve the capture of global context. In contrast, NARM relies on local temporal patterns and can be more sensitive to substitution-based perturbations.

\section{Conclusion and Future Work}
In this work, we study the effectiveness of substitution-based profile pollution attacks on sequential recommenders. We design a profile pollution attack algorithm and evaluate on multiple datasets and state-of-the-art recommenders, where the proposed method successfully poses threats to sequential recommenders for both the untargeted and targeted attack scenarios. Then, we explore the defense methods by sampling augmented item representations from a Dirichlet distribution within multi-hop neighborhoods. We additionally design an adversarial training method to search for the worst-case augmentation and enhance model robustness. To the best of our knowledge, this is the first work on defending substitution-based profile pollution attacks for sequential recommenders. Our experimental results suggest that by applying the proposed defense methods, sequential recommender systems can learn robust item representations and demonstrate significantly reduced performance variations under profile pollution attacks.

The proposed methods have certain limitations. For example, our attack and defense methods are designed for sequential recommenders and might not be applicable for other recommender systems. Additionally, the profile pollution experiments are performed under a white-box assumption, rendering the proposed method less accessible in real-world scenarios. Despite having introduced Dirichlet neighborhood sampling, we have not discussed different choices of $\alpha$ values for multi-hop neighbors to exploit the potential benefits of the Dirichlet distribution.
For future work, we plan to relax our assumptions and explore certified defense methods for sequential recommenders. For example, it would be interesting to explore interval bound propagation to provide guaranteed defense. Additionally, we can design model-agnostic adversarial attack and defense algorithms to extend our framework to other architectures.


\begin{acks}
This research is supported in part by the National Science Foundation under Grant No. IIS-2202481, CHE-2105032,  IIS-2008228, CNS-1845639, CNS-1831669. The views and conclusions contained in this document are those of the authors and should not be interpreted as representing the official policies, either expressed or implied, of the U.S. Government. The U.S. Government is authorized to reproduce and distribute reprints for Government purposes notwithstanding any copyright notation here on.
\end{acks}

\bibliographystyle{ACM-Reference-Format}
\bibliography{reference}


\begin{thebibliography}{45}


\ifx \showCODEN    \undefined \def \showCODEN     #1{\unskip}     \fi
\ifx \showDOI      \undefined \def \showDOI       #1{#1}\fi
\ifx \showISBNx    \undefined \def \showISBNx     #1{\unskip}     \fi
\ifx \showISBNxiii \undefined \def \showISBNxiii  #1{\unskip}     \fi
\ifx \showISSN     \undefined \def \showISSN      #1{\unskip}     \fi
\ifx \showLCCN     \undefined \def \showLCCN      #1{\unskip}     \fi
\ifx \shownote     \undefined \def \shownote      #1{#1}          \fi
\ifx \showarticletitle \undefined \def \showarticletitle #1{#1}   \fi
\ifx \showURL      \undefined \def \showURL       {\relax}        \fi
\providecommand\bibfield[2]{#2}
\providecommand\bibinfo[2]{#2}
\providecommand\natexlab[1]{#1}
\providecommand\showeprint[2][]{arXiv:#2}

\bibitem[Alzantot et~al\mbox{.}(2018)]%
        {alzantot2018generating}
\bibfield{author}{\bibinfo{person}{Moustafa Alzantot}, \bibinfo{person}{Yash
  Sharma}, \bibinfo{person}{Ahmed Elgohary}, \bibinfo{person}{Bo-Jhang Ho},
  \bibinfo{person}{Mani Srivastava}, {and} \bibinfo{person}{Kai-Wei Chang}.}
  \bibinfo{year}{2018}\natexlab{}.
\newblock \showarticletitle{Generating natural language adversarial examples}.
\newblock \bibinfo{journal}{\emph{arXiv preprint arXiv:1804.07998}}
  (\bibinfo{year}{2018}).
\newblock


\bibitem[Celma(2010)]%
        {Celma:Springer2010}
\bibfield{author}{\bibinfo{person}{O. Celma}.} \bibinfo{year}{2010}\natexlab{}.
\newblock \bibinfo{booktitle}{\emph{{Music Recommendation and Discovery in the
  Long Tail}}}.
\newblock \bibinfo{publisher}{Springer}.
\newblock


\bibitem[Chen and Li(2019)]%
        {chen2019adversarial}
\bibfield{author}{\bibinfo{person}{Huiyuan Chen} {and} \bibinfo{person}{Jing
  Li}.} \bibinfo{year}{2019}\natexlab{}.
\newblock \showarticletitle{Adversarial tensor factorization for context-aware
  recommendation}. In \bibinfo{booktitle}{\emph{Proceedings of the 13th ACM
  Conference on Recommender Systems}}. \bibinfo{pages}{363--367}.
\newblock


\bibitem[Christakopoulou and Banerjee(2019)]%
        {christakopoulou2019adversarial}
\bibfield{author}{\bibinfo{person}{Konstantina Christakopoulou} {and}
  \bibinfo{person}{Arindam Banerjee}.} \bibinfo{year}{2019}\natexlab{}.
\newblock \showarticletitle{Adversarial attacks on an oblivious recommender}.
  In \bibinfo{booktitle}{\emph{Proceedings of the 13th ACM Conference on
  Recommender Systems}}. \bibinfo{pages}{322--330}.
\newblock


\bibitem[Ebrahimi et~al\mbox{.}(2017)]%
        {ebrahimi2017hotflip}
\bibfield{author}{\bibinfo{person}{Javid Ebrahimi}, \bibinfo{person}{Anyi Rao},
  \bibinfo{person}{Daniel Lowd}, {and} \bibinfo{person}{Dejing Dou}.}
  \bibinfo{year}{2017}\natexlab{}.
\newblock \showarticletitle{Hotflip: White-box adversarial examples for text
  classification}.
\newblock \bibinfo{journal}{\emph{arXiv preprint arXiv:1712.06751}}
  (\bibinfo{year}{2017}).
\newblock


\bibitem[Fang et~al\mbox{.}(2020)]%
        {fang2020influence}
\bibfield{author}{\bibinfo{person}{Minghong Fang},
  \bibinfo{person}{Neil~Zhenqiang Gong}, {and} \bibinfo{person}{Jia Liu}.}
  \bibinfo{year}{2020}\natexlab{}.
\newblock \showarticletitle{Influence function based data poisoning attacks to
  top-n recommender systems}. In \bibinfo{booktitle}{\emph{Proceedings of The
  Web Conference 2020}}. \bibinfo{pages}{3019--3025}.
\newblock


\bibitem[Garg and Ramakrishnan(2020)]%
        {garg2020bae}
\bibfield{author}{\bibinfo{person}{Siddhant Garg} {and}
  \bibinfo{person}{Goutham Ramakrishnan}.} \bibinfo{year}{2020}\natexlab{}.
\newblock \showarticletitle{Bae: Bert-based adversarial examples for text
  classification}.
\newblock \bibinfo{journal}{\emph{arXiv preprint arXiv:2004.01970}}
  (\bibinfo{year}{2020}).
\newblock


\bibitem[Geirhos et~al\mbox{.}(2020)]%
        {geirhos2020shortcut}
\bibfield{author}{\bibinfo{person}{Robert Geirhos},
  \bibinfo{person}{J{\"o}rn-Henrik Jacobsen}, \bibinfo{person}{Claudio
  Michaelis}, \bibinfo{person}{Richard Zemel}, \bibinfo{person}{Wieland
  Brendel}, \bibinfo{person}{Matthias Bethge}, {and} \bibinfo{person}{Felix~A
  Wichmann}.} \bibinfo{year}{2020}\natexlab{}.
\newblock \showarticletitle{Shortcut learning in deep neural networks}.
\newblock \bibinfo{journal}{\emph{Nature Machine Intelligence}}
  \bibinfo{volume}{2}, \bibinfo{number}{11} (\bibinfo{year}{2020}),
  \bibinfo{pages}{665--673}.
\newblock


\bibitem[Goodfellow et~al\mbox{.}(2014)]%
        {goodfellow2014explaining}
\bibfield{author}{\bibinfo{person}{Ian~J Goodfellow}, \bibinfo{person}{Jonathon
  Shlens}, {and} \bibinfo{person}{Christian Szegedy}.}
  \bibinfo{year}{2014}\natexlab{}.
\newblock \showarticletitle{Explaining and harnessing adversarial examples}.
\newblock \bibinfo{journal}{\emph{arXiv preprint arXiv:1412.6572}}
  (\bibinfo{year}{2014}).
\newblock


\bibitem[Gowal et~al\mbox{.}(2018)]%
        {gowal2018effectiveness}
\bibfield{author}{\bibinfo{person}{Sven Gowal}, \bibinfo{person}{Krishnamurthy
  Dvijotham}, \bibinfo{person}{Robert Stanforth}, \bibinfo{person}{Rudy Bunel},
  \bibinfo{person}{Chongli Qin}, \bibinfo{person}{Jonathan Uesato},
  \bibinfo{person}{Relja Arandjelovic}, \bibinfo{person}{Timothy Mann}, {and}
  \bibinfo{person}{Pushmeet Kohli}.} \bibinfo{year}{2018}\natexlab{}.
\newblock \showarticletitle{On the effectiveness of interval bound propagation
  for training verifiably robust models}.
\newblock \bibinfo{journal}{\emph{arXiv preprint arXiv:1810.12715}}
  (\bibinfo{year}{2018}).
\newblock


\bibitem[Harper and Konstan(2015)]%
        {harper2015movielens}
\bibfield{author}{\bibinfo{person}{F~Maxwell Harper} {and}
  \bibinfo{person}{Joseph~A Konstan}.} \bibinfo{year}{2015}\natexlab{}.
\newblock \showarticletitle{The movielens datasets: History and context}.
\newblock \bibinfo{journal}{\emph{Acm transactions on interactive intelligent
  systems (tiis)}} \bibinfo{volume}{5}, \bibinfo{number}{4}
  (\bibinfo{year}{2015}), \bibinfo{pages}{1--19}.
\newblock


\bibitem[He et~al\mbox{.}(2018)]%
        {he2018adversarial}
\bibfield{author}{\bibinfo{person}{Xiangnan He}, \bibinfo{person}{Zhankui He},
  \bibinfo{person}{Xiaoyu Du}, {and} \bibinfo{person}{Tat-Seng Chua}.}
  \bibinfo{year}{2018}\natexlab{}.
\newblock \showarticletitle{Adversarial personalized ranking for
  recommendation}. In \bibinfo{booktitle}{\emph{The 41st International ACM
  SIGIR Conference on Research \& Development in Information Retrieval}}.
  \bibinfo{pages}{355--364}.
\newblock


\bibitem[He et~al\mbox{.}(2021)]%
        {he2021locker}
\bibfield{author}{\bibinfo{person}{Zhankui He}, \bibinfo{person}{Handong Zhao},
  \bibinfo{person}{Zhe Lin}, \bibinfo{person}{Zhaowen Wang},
  \bibinfo{person}{Ajinkya Kale}, {and} \bibinfo{person}{Julian Mcauley}.}
  \bibinfo{year}{2021}\natexlab{}.
\newblock \showarticletitle{Locker: Locally Constrained Self-Attentive
  Sequential Recommendation}. In \bibinfo{booktitle}{\emph{Proceedings of the
  30th ACM International Conference on Information \& Knowledge Management}}.
  \bibinfo{pages}{3088--3092}.
\newblock


\bibitem[Huang et~al\mbox{.}(2021)]%
        {huang2021data}
\bibfield{author}{\bibinfo{person}{Hai Huang}, \bibinfo{person}{Jiaming Mu},
  \bibinfo{person}{Neil~Zhenqiang Gong}, \bibinfo{person}{Qi Li},
  \bibinfo{person}{Bin Liu}, {and} \bibinfo{person}{Mingwei Xu}.}
  \bibinfo{year}{2021}\natexlab{}.
\newblock \showarticletitle{Data Poisoning Attacks to Deep Learning Based
  Recommender Systems}.
\newblock \bibinfo{journal}{\emph{arXiv preprint arXiv:2101.02644}}
  (\bibinfo{year}{2021}).
\newblock


\bibitem[Kang and McAuley(2018)]%
        {kang2018self}
\bibfield{author}{\bibinfo{person}{Wang-Cheng Kang} {and}
  \bibinfo{person}{Julian McAuley}.} \bibinfo{year}{2018}\natexlab{}.
\newblock \showarticletitle{Self-attentive sequential recommendation}. In
  \bibinfo{booktitle}{\emph{2018 IEEE International Conference on Data Mining
  (ICDM)}}. IEEE, \bibinfo{pages}{197--206}.
\newblock


\bibitem[Lee et~al\mbox{.}(2017)]%
        {lee2017all}
\bibfield{author}{\bibinfo{person}{Sungho Lee}, \bibinfo{person}{Sungjae
  Hwang}, {and} \bibinfo{person}{Sukyoung Ryu}.}
  \bibinfo{year}{2017}\natexlab{}.
\newblock \showarticletitle{All about activity injection: Threats, semantics,
  and detection}. In \bibinfo{booktitle}{\emph{2017 32nd IEEE/ACM International
  Conference on Automated Software Engineering (ASE)}}. IEEE,
  \bibinfo{pages}{252--262}.
\newblock


\bibitem[Li et~al\mbox{.}(2018)]%
        {li2018textbugger}
\bibfield{author}{\bibinfo{person}{Jinfeng Li}, \bibinfo{person}{Shouling Ji},
  \bibinfo{person}{Tianyu Du}, \bibinfo{person}{Bo Li}, {and}
  \bibinfo{person}{Ting Wang}.} \bibinfo{year}{2018}\natexlab{}.
\newblock \showarticletitle{Textbugger: Generating adversarial text against
  real-world applications}.
\newblock \bibinfo{journal}{\emph{arXiv preprint arXiv:1812.05271}}
  (\bibinfo{year}{2018}).
\newblock


\bibitem[Li et~al\mbox{.}(2017)]%
        {li2017neural}
\bibfield{author}{\bibinfo{person}{Jing Li}, \bibinfo{person}{Pengjie Ren},
  \bibinfo{person}{Zhumin Chen}, \bibinfo{person}{Zhaochun Ren},
  \bibinfo{person}{Tao Lian}, {and} \bibinfo{person}{Jun Ma}.}
  \bibinfo{year}{2017}\natexlab{}.
\newblock \showarticletitle{Neural attentive session-based recommendation}. In
  \bibinfo{booktitle}{\emph{Proceedings of the 2017 ACM on Conference on
  Information and Knowledge Management}}. \bibinfo{pages}{1419--1428}.
\newblock


\bibitem[Li et~al\mbox{.}(2020)]%
        {li2020bert}
\bibfield{author}{\bibinfo{person}{Linyang Li}, \bibinfo{person}{Ruotian Ma},
  \bibinfo{person}{Qipeng Guo}, \bibinfo{person}{Xiangyang Xue}, {and}
  \bibinfo{person}{Xipeng Qiu}.} \bibinfo{year}{2020}\natexlab{}.
\newblock \showarticletitle{Bert-attack: Adversarial attack against bert using
  bert}.
\newblock \bibinfo{journal}{\emph{arXiv preprint arXiv:2004.09984}}
  (\bibinfo{year}{2020}).
\newblock


\bibitem[Marshall and Wang(2016)]%
        {marshall2016mood}
\bibfield{author}{\bibinfo{person}{Jermaine Marshall} {and}
  \bibinfo{person}{Dong Wang}.} \bibinfo{year}{2016}\natexlab{}.
\newblock \showarticletitle{Mood-sensitive truth discovery for reliable
  recommendation systems in social sensing}. In
  \bibinfo{booktitle}{\emph{Proceedings of the 10th ACM conference on
  recommender systems}}. \bibinfo{pages}{167--174}.
\newblock


\bibitem[McAuley et~al\mbox{.}(2015)]%
        {mcauley2015image}
\bibfield{author}{\bibinfo{person}{Julian McAuley},
  \bibinfo{person}{Christopher Targett}, \bibinfo{person}{Qinfeng Shi}, {and}
  \bibinfo{person}{Anton Van Den~Hengel}.} \bibinfo{year}{2015}\natexlab{}.
\newblock \showarticletitle{Image-based recommendations on styles and
  substitutes}. In \bibinfo{booktitle}{\emph{Proceedings of the 38th
  international ACM SIGIR conference on research and development in information
  retrieval}}. \bibinfo{pages}{43--52}.
\newblock


\bibitem[Meng et~al\mbox{.}(2014)]%
        {meng2014your}
\bibfield{author}{\bibinfo{person}{Wei Meng}, \bibinfo{person}{Xinyu Xing},
  \bibinfo{person}{Anmol Sheth}, \bibinfo{person}{Udi Weinsberg}, {and}
  \bibinfo{person}{Wenke Lee}.} \bibinfo{year}{2014}\natexlab{}.
\newblock \showarticletitle{Your online interests: Pwned! a pollution attack
  against targeted advertising}. In \bibinfo{booktitle}{\emph{Proceedings of
  the 2014 ACM SIGSAC Conference on Computer and Communications Security}}.
  \bibinfo{pages}{129--140}.
\newblock


\bibitem[Ni et~al\mbox{.}(2019)]%
        {ni-etal-2019-justifying}
\bibfield{author}{\bibinfo{person}{Jianmo Ni}, \bibinfo{person}{Jiacheng Li},
  {and} \bibinfo{person}{Julian McAuley}.} \bibinfo{year}{2019}\natexlab{}.
\newblock \showarticletitle{Justifying Recommendations using Distantly-Labeled
  Reviews and Fine-Grained Aspects}. In \bibinfo{booktitle}{\emph{Proceedings
  of the 2019 Conference on Empirical Methods in Natural Language Processing
  and the 9th International Joint Conference on Natural Language Processing
  (EMNLP-IJCNLP)}}. \bibinfo{publisher}{Association for Computational
  Linguistics}, \bibinfo{address}{Hong Kong, China}, \bibinfo{pages}{188--197}.
\newblock
\urldef\tempurl%
\url{https://doi.org/10.18653/v1/D19-1018}
\showDOI{\tempurl}


\bibitem[Oh and Kumar(2022)]%
        {oh2022robustness}
\bibfield{author}{\bibinfo{person}{Sejoon Oh} {and} \bibinfo{person}{Srijan
  Kumar}.} \bibinfo{year}{2022}\natexlab{}.
\newblock \showarticletitle{Robustness of Deep Recommendation Systems to
  Untargeted Interaction Perturbations}.
\newblock \bibinfo{journal}{\emph{arXiv preprint arXiv:2201.12686}}
  (\bibinfo{year}{2022}).
\newblock


\bibitem[O'Mahony et~al\mbox{.}(2005)]%
        {o2005recommender}
\bibfield{author}{\bibinfo{person}{Michael~P O'Mahony}, \bibinfo{person}{Neil~J
  Hurley}, {and} \bibinfo{person}{Gu{\'e}nol{\'e}~CM Silvestre}.}
  \bibinfo{year}{2005}\natexlab{}.
\newblock \showarticletitle{Recommender systems: Attack types and strategies}.
  In \bibinfo{booktitle}{\emph{AAAI}}. \bibinfo{pages}{334--339}.
\newblock


\bibitem[Ren et~al\mbox{.}(2015)]%
        {ren2015towards}
\bibfield{author}{\bibinfo{person}{Chuangang Ren}, \bibinfo{person}{Yulong
  Zhang}, \bibinfo{person}{Hui Xue}, \bibinfo{person}{Tao Wei}, {and}
  \bibinfo{person}{Peng Liu}.} \bibinfo{year}{2015}\natexlab{}.
\newblock \showarticletitle{Towards discovering and understanding task
  hijacking in android}. In \bibinfo{booktitle}{\emph{24th $\{$USENIX$\}$
  Security Symposium ($\{$USENIX$\}$ Security 15)}}. \bibinfo{pages}{945--959}.
\newblock


\bibitem[Ren et~al\mbox{.}(2019)]%
        {ren2019generating}
\bibfield{author}{\bibinfo{person}{Shuhuai Ren}, \bibinfo{person}{Yihe Deng},
  \bibinfo{person}{Kun He}, {and} \bibinfo{person}{Wanxiang Che}.}
  \bibinfo{year}{2019}\natexlab{}.
\newblock \showarticletitle{Generating natural language adversarial examples
  through probability weighted word saliency}. In
  \bibinfo{booktitle}{\emph{Proceedings of the 57th annual meeting of the
  association for computational linguistics}}. \bibinfo{pages}{1085--1097}.
\newblock


\bibitem[Sato et~al\mbox{.}(2018)]%
        {sato2018interpretable}
\bibfield{author}{\bibinfo{person}{Motoki Sato}, \bibinfo{person}{Jun Suzuki},
  \bibinfo{person}{Hiroyuki Shindo}, {and} \bibinfo{person}{Yuji Matsumoto}.}
  \bibinfo{year}{2018}\natexlab{}.
\newblock \showarticletitle{Interpretable adversarial perturbation in input
  embedding space for text}.
\newblock \bibinfo{journal}{\emph{arXiv preprint arXiv:1805.02917}}
  (\bibinfo{year}{2018}).
\newblock


\bibitem[Song et~al\mbox{.}(2020)]%
        {song2020poisonrec}
\bibfield{author}{\bibinfo{person}{Junshuai Song}, \bibinfo{person}{Zhao Li},
  \bibinfo{person}{Zehong Hu}, \bibinfo{person}{Yucheng Wu},
  \bibinfo{person}{Zhenpeng Li}, \bibinfo{person}{Jian Li}, {and}
  \bibinfo{person}{Jun Gao}.} \bibinfo{year}{2020}\natexlab{}.
\newblock \showarticletitle{Poisonrec: an adaptive data poisoning framework for
  attacking black-box recommender systems}. In \bibinfo{booktitle}{\emph{2020
  IEEE 36th International Conference on Data Engineering (ICDE)}}. IEEE,
  \bibinfo{pages}{157--168}.
\newblock


\bibitem[Sun et~al\mbox{.}(2019)]%
        {sun2019bert4rec}
\bibfield{author}{\bibinfo{person}{Fei Sun}, \bibinfo{person}{Jun Liu},
  \bibinfo{person}{Jian Wu}, \bibinfo{person}{Changhua Pei},
  \bibinfo{person}{Xiao Lin}, \bibinfo{person}{Wenwu Ou}, {and}
  \bibinfo{person}{Peng Jiang}.} \bibinfo{year}{2019}\natexlab{}.
\newblock \showarticletitle{BERT4Rec: Sequential recommendation with
  bidirectional encoder representations from transformer}. In
  \bibinfo{booktitle}{\emph{Proceedings of the 28th ACM International
  Conference on Information and Knowledge Management}}.
  \bibinfo{pages}{1441--1450}.
\newblock


\bibitem[Tang et~al\mbox{.}(2019)]%
        {tang2019adversarial}
\bibfield{author}{\bibinfo{person}{Jinhui Tang}, \bibinfo{person}{Xiaoyu Du},
  \bibinfo{person}{Xiangnan He}, \bibinfo{person}{Fajie Yuan},
  \bibinfo{person}{Qi Tian}, {and} \bibinfo{person}{Tat-Seng Chua}.}
  \bibinfo{year}{2019}\natexlab{}.
\newblock \showarticletitle{Adversarial training towards robust multimedia
  recommender system}.
\newblock \bibinfo{journal}{\emph{IEEE Transactions on Knowledge and Data
  Engineering}} \bibinfo{volume}{32}, \bibinfo{number}{5}
  (\bibinfo{year}{2019}), \bibinfo{pages}{855--867}.
\newblock


\bibitem[Tang et~al\mbox{.}(2020)]%
        {tang2020revisiting}
\bibfield{author}{\bibinfo{person}{Jiaxi Tang}, \bibinfo{person}{Hongyi Wen},
  {and} \bibinfo{person}{Ke Wang}.} \bibinfo{year}{2020}\natexlab{}.
\newblock \showarticletitle{Revisiting Adversarially Learned Injection Attacks
  Against Recommender Systems}. In \bibinfo{booktitle}{\emph{Fourteenth ACM
  Conference on Recommender Systems}}. \bibinfo{pages}{318--327}.
\newblock


\bibitem[Xing et~al\mbox{.}(2013)]%
        {xing2013take}
\bibfield{author}{\bibinfo{person}{Xingyu Xing}, \bibinfo{person}{Wei Meng},
  \bibinfo{person}{Dan Doozan}, \bibinfo{person}{Alex~C Snoeren},
  \bibinfo{person}{Nick Feamster}, {and} \bibinfo{person}{Wenke Lee}.}
  \bibinfo{year}{2013}\natexlab{}.
\newblock \showarticletitle{Take this personally: Pollution attacks on
  personalized services}. In \bibinfo{booktitle}{\emph{22nd USENIX Security
  Symposium (USENIX Security 13)}}. \bibinfo{pages}{671--686}.
\newblock


\bibitem[Yang et~al\mbox{.}(2017)]%
        {yang2017fake}
\bibfield{author}{\bibinfo{person}{Guolei Yang},
  \bibinfo{person}{Neil~Zhenqiang Gong}, {and} \bibinfo{person}{Ying Cai}.}
  \bibinfo{year}{2017}\natexlab{}.
\newblock \showarticletitle{Fake Co-visitation Injection Attacks to Recommender
  Systems.}. In \bibinfo{booktitle}{\emph{NDSS}}.
\newblock


\bibitem[Yuan et~al\mbox{.}(2019)]%
        {yuan2019adversarial}
\bibfield{author}{\bibinfo{person}{Feng Yuan}, \bibinfo{person}{Lina Yao},
  {and} \bibinfo{person}{Boualem Benatallah}.} \bibinfo{year}{2019}\natexlab{}.
\newblock \showarticletitle{Adversarial collaborative neural network for robust
  recommendation}. In \bibinfo{booktitle}{\emph{Proceedings of the 42nd
  International ACM SIGIR Conference on Research and Development in Information
  Retrieval}}. \bibinfo{pages}{1065--1068}.
\newblock


\bibitem[Yue et~al\mbox{.}(2021)]%
        {yue2021black}
\bibfield{author}{\bibinfo{person}{Zhenrui Yue}, \bibinfo{person}{Zhankui He},
  \bibinfo{person}{Huimin Zeng}, {and} \bibinfo{person}{Julian McAuley}.}
  \bibinfo{year}{2021}\natexlab{}.
\newblock \showarticletitle{Black-Box Attacks on Sequential Recommenders via
  Data-Free Model Extraction}. In \bibinfo{booktitle}{\emph{Fifteenth ACM
  Conference on Recommender Systems}}. \bibinfo{pages}{44--54}.
\newblock


\bibitem[Zeller and Felten(2008)]%
        {zeller2008cross}
\bibfield{author}{\bibinfo{person}{William Zeller} {and}
  \bibinfo{person}{Edward~W Felten}.} \bibinfo{year}{2008}\natexlab{}.
\newblock \showarticletitle{Cross-site request forgeries: Exploitation and
  prevention}.
\newblock \bibinfo{journal}{\emph{Bericht, Princeton University}}
  (\bibinfo{year}{2008}).
\newblock


\bibitem[Zeng et~al\mbox{.}(2021a)]%
        {zeng2021certified}
\bibfield{author}{\bibinfo{person}{Huimin Zeng}, \bibinfo{person}{Jiahao Su},
  {and} \bibinfo{person}{Furong Huang}.} \bibinfo{year}{2021}\natexlab{a}.
\newblock \showarticletitle{Certified Defense via Latent Space Randomized
  Smoothing with Orthogonal Encoders}.
\newblock \bibinfo{journal}{\emph{arXiv preprint arXiv:2108.00491}}
  (\bibinfo{year}{2021}).
\newblock


\bibitem[Zeng et~al\mbox{.}(2021b)]%
        {zeng2021adversarial}
\bibfield{author}{\bibinfo{person}{Huimin Zeng}, \bibinfo{person}{Chen Zhu},
  \bibinfo{person}{Tom Goldstein}, {and} \bibinfo{person}{Furong Huang}.}
  \bibinfo{year}{2021}\natexlab{b}.
\newblock \showarticletitle{Are adversarial examples created equal? A learnable
  weighted minimax risk for robustness under non-uniform attacks}. In
  \bibinfo{booktitle}{\emph{Proceedings of the AAAI Conference on Artificial
  Intelligence}}, Vol.~\bibinfo{volume}{35}. \bibinfo{pages}{10815--10823}.
\newblock


\bibitem[Zhang et~al\mbox{.}(2019a)]%
        {zhang2019through}
\bibfield{author}{\bibinfo{person}{Daniel Zhang}, \bibinfo{person}{Bo Ni},
  \bibinfo{person}{Qiyu Zhi}, \bibinfo{person}{Thomas Plummer},
  \bibinfo{person}{Qi Li}, \bibinfo{person}{Hao Zheng},
  \bibinfo{person}{Qingkai Zeng}, \bibinfo{person}{Yang Zhang}, {and}
  \bibinfo{person}{Dong Wang}.} \bibinfo{year}{2019}\natexlab{a}.
\newblock \showarticletitle{Through the eyes of a poet: Classical poetry
  recommendation with visual input on social media}. In
  \bibinfo{booktitle}{\emph{2019 IEEE/ACM International Conference on Advances
  in Social Networks Analysis and Mining (ASONAM)}}. IEEE,
  \bibinfo{pages}{333--340}.
\newblock


\bibitem[Zhang et~al\mbox{.}(2018)]%
        {zhang2018crowdsourcing}
\bibfield{author}{\bibinfo{person}{Daniel~Yue Zhang}, \bibinfo{person}{Qi Li},
  \bibinfo{person}{Herman Tong}, \bibinfo{person}{Jose Badilla},
  \bibinfo{person}{Yang Zhang}, {and} \bibinfo{person}{Dong Wang}.}
  \bibinfo{year}{2018}\natexlab{}.
\newblock \showarticletitle{Crowdsourcing-based copyright infringement
  detection in live video streams}. In \bibinfo{booktitle}{\emph{2018 IEEE/ACM
  International Conference on Advances in Social Networks Analysis and Mining
  (ASONAM)}}. IEEE, \bibinfo{pages}{367--374}.
\newblock


\bibitem[Zhang et~al\mbox{.}(2020)]%
        {zhang2020practical}
\bibfield{author}{\bibinfo{person}{Hengtong Zhang}, \bibinfo{person}{Yaliang
  Li}, \bibinfo{person}{Bolin Ding}, {and} \bibinfo{person}{Jing Gao}.}
  \bibinfo{year}{2020}\natexlab{}.
\newblock \showarticletitle{Practical data poisoning attack against next-item
  recommendation}. In \bibinfo{booktitle}{\emph{Proceedings of The Web
  Conference 2020}}. \bibinfo{pages}{2458--2464}.
\newblock


\bibitem[Zhang et~al\mbox{.}(2019b)]%
        {zhang2019understanding}
\bibfield{author}{\bibinfo{person}{Yubao Zhang}, \bibinfo{person}{Jidong Xiao},
  \bibinfo{person}{Shuai Hao}, \bibinfo{person}{Haining Wang},
  \bibinfo{person}{Sencun Zhu}, {and} \bibinfo{person}{Sushil Jajodia}.}
  \bibinfo{year}{2019}\natexlab{b}.
\newblock \showarticletitle{Understanding the manipulation on recommender
  systems through web injection}.
\newblock \bibinfo{journal}{\emph{IEEE Transactions on Information Forensics
  and Security}}  \bibinfo{volume}{15} (\bibinfo{year}{2019}),
  \bibinfo{pages}{3807--3818}.
\newblock


\bibitem[Zhou et~al\mbox{.}(2020)]%
        {zhou2020defense}
\bibfield{author}{\bibinfo{person}{Yi Zhou}, \bibinfo{person}{Xiaoqing Zheng},
  \bibinfo{person}{Cho-Jui Hsieh}, \bibinfo{person}{Kai-wei Chang}, {and}
  \bibinfo{person}{Xuanjing Huang}.} \bibinfo{year}{2020}\natexlab{}.
\newblock \showarticletitle{Defense against adversarial attacks in nlp via
  dirichlet neighborhood ensemble}.
\newblock \bibinfo{journal}{\emph{arXiv preprint arXiv:2006.11627}}
  (\bibinfo{year}{2020}).
\newblock


\bibitem[Zhu et~al\mbox{.}(2019)]%
        {zhu2019freelb}
\bibfield{author}{\bibinfo{person}{Chen Zhu}, \bibinfo{person}{Yu Cheng},
  \bibinfo{person}{Zhe Gan}, \bibinfo{person}{Siqi Sun}, \bibinfo{person}{Tom
  Goldstein}, {and} \bibinfo{person}{Jingjing Liu}.}
  \bibinfo{year}{2019}\natexlab{}.
\newblock \showarticletitle{Freelb: Enhanced adversarial training for language
  understanding}.
\newblock  (\bibinfo{year}{2019}).
\newblock


\end{thebibliography}

\end{document}